\documentclass[preprint,showpacs,preprintnumbers,amsmath,amssymb]{revtex4}

\usepackage{graphicx}% Include figure files
\usepackage{dcolumn}% Align table columns on decimal point
\usepackage{bm}% bold math

\begin{document}

%\preprint{APS/123-QED}

\title{Multi-mode theory of pulsed twin beams generation using a high gain fiber optical parametric amplifier}

%\title{Quantum noise of a high gain fiber optical parametric amplifier using pulsed pump }% Force line breaks with \\

\author{Xueshi Guo$^{\text{1}}$, Xiaoying Li$^{\text{1},*}$\footnotetext{* Email: xiaoyingli@tju.edu.cn}, Nannan Liu$^{\text{1}}$, Z. Y. Ou$^{\text{2},\dag}$ \footnotetext{\dag Email: zou@iupui.edu}}

\affiliation{$^{\text{1}}$ College of Precision Instrument and
Opto-electronics Engineering, Tianjin University, \\Key Laboratory of Optoelectronics Information Technology, Ministry of Education, Tianjin, 300072,
P. R. China\\$^{\text{2}}$ Department of Physics, Indiana University-Purdue University Indianapolis, Indianapolis, IN 46202, USA}

%\address{$^{*}$Corresponding author: xiaoyingli@tju.edu.cn}

\date{\today}

\begin{abstract}
We theoretically investigate the quantum noise properties of the pulse pumped high gain fiber optical parametric amplifiers (FOPA) by using the Bogoliubov transformation in multi-frequency modes to describe the evolution of the non-degenerate signal and idler twin beams. The results show that the noise figure of the FOPA is generally greater than the 3 dB quantum limit unless the joint spectral function is factorable and the spectrum of the input signal well matches the gain spectrum in the signal band. However, the intensity difference noise of the twin beams, which weakly depends on the joint spectral function, can be significantly less than the shot-noise limit when the temporal modes of the pump and the input signal are properly matched. Moreover, to closely resemble the real experimental condition, the quantum noise of twin beams generated from a broadband FOPA is numerically studied by taking the various kinds of experimental imperfections into account. Our study is not only useful for developing a compact fiber source of twin beams, but also helpful for understanding the quantum noise limit of a pulse pumped FOPA in the fiber communication system.

%\begin{description}
%\end{description}
\end{abstract}

\pacs{42.50.Dv, 42.65.Yj, 03.67.Hk}% PACS, the Physics and Astronomy
\keywords{Quantum Optics, Parametric amplifier, Four-wave Mixing}%Use showkeys class option if keyword
\maketitle

%\tableofcontents
%\reference on nature photonic is not correct

\section{\label{INTRODUCTION}Introduction:\protect\\}

It is well known that the fiber optical parametric amplifiers (FOPAs), employing $\chi^{(3)}$ nonlinearity based four wave mixing (FWM) to transfer energy from one or two strong pump fields to a weak
signal field, can be used in the conventional optical communication systems for
signal-processing applications, such as optical amplification, phase conjugation, optical limiters, 3R generators, and time-domain de-multiplexer etc.~\cite{Tong2011,Hans02,Marhic,Agrawal08}. In fact, similar to the $\chi^{(2)}$ crystals based optical parametric amplifiers (OPAs)~\cite{Caves1982A,wu86,Ayt90,Reid09}, FOPAs are also candidates for generating non-classical light for quantum communication~\cite{Sharping01,Voss06,Mck12}.

In quantum communication, information can be encoded discretely on each photon or continuously in quadrature-phase amplitudes of an optical field. The former requires single-photon sources that can be generated with weak nonlinear interaction~\cite{Tittel01}, but the latter relies on strong interaction to generate quantum correlation between the amplitudes of optical fields for entanglement~\cite{Braun05,Reid09}. Moreover, continuous variable (CV) entanglement offers several advantages over its discrete variable
counterpart. The main advantage is the experimental feasibility in unconditional production of entangled states which in turn enables quantum information protocols to be accomplished unconditionally~\cite{Braun05}.

The past decade has seen the growing interest in developing nonclassical light sources via the FWM in fibers, because they are compatible with the fiber network~\cite{Fiorentino02,Li05a,Takesue06,Alibart06,Fan08}. In order to enhance the FWM and to suppress the Raman scattering~\cite{Li04}, a pulse pumped FOPA is preferred because the peak power of pulsed laser is very high even at a modest average power. However, for the majority of the experiments performed to date, the FWM in fiber is in the low gain regime, and the fiber based nonclassical light sources with vacuum injection are in the domain of a few photons and thus can be only employed for discrete variable information encoding. In order to develop the fiber based source of the CV nonclassical light, the pulse pumped FOPA needs to be operated in the high gain regime.

Up to now, only a couple of experiments reported the generation of CV nonclassical light via the pulse pumped FWM in fibers \cite{Sharping01, Guo12}. In 2001, Sharping and co-works realized the pulsed twin beams by using the FOPA for the first time \cite{Sharping01}. The amplified signal and generated idler beams bear a strong quantum correlation in intensity (photon-number) fluctuations in the sense that the quantum noise level
of their intensity difference is lower than the shot-noise limit (SNL). In that experiment, the gain of FWM
is less than 3 and the observed intensity difference noise of twin beams is only about 1.1 dB (2.6 dB after correction for losses) lower than the SNL. Recently, using the FOPA with photon number gain of about 16 dB, our group demonstrated the twin beams with
the intensity-difference noise below the SNL by 3.1 dB (10.4 dB after correction for losses) \cite{Guo12}. The results indicate that compared with its $\chi^{(2)}$ crystal counterparts~\cite{Ayt90,Lau03,Zhang07}, the pulse pumped high gain FOPA is a simple and promising system for generating the CV nonclassical light, which has the potential applications in realizing the quantum communication protocols and in performing the high precision measurement.

There have been a few theoretical papers analyzing the generation of pulsed CV nonclassical light via the high gain $\chi^{(2)}$ crystals based OPAs~\cite{Wasi06,Silberhorn2011}. For the pulse pumped high gain FOPAs, however, the investigation is mostly within the domain of classical characteristics~\cite{Hans02,Marhic} and the quantum theory has not been worked out yet. This is different from the case of continuous wave (CW) or quasi-CW pumped high gain FOPAs~\cite{Hans02,Voss06,Mck12}. In order to understand the high gain FOPA for further reducing the intensity difference noise of pulsed twin beams, we focus on developing a multi-mode quantum theory in this paper.

The rest of the paper is organized as follows. After briefly introducing the quantum description of a single-mode FOPA in Sec. II, we deduce a multi-mode quantum theory of a pulse pumped non-degenerate FOPA in Sec. III. The evolution of non-degenerate signal and idler beams is described by
the Bogoliubov transformation in a multi-frequency mode form. According to the Hamiltonian of FWM in fibers, we derive the four transformation functions of the Bogoliubov transformation in the analytical form of infinite series. The calculation shows that the noise performance of the high gain FOPA highly depends on the joint spectral function (JSF). In Sec. IV, we show that the FOPA with factorized JSF can be described by a quantum model of single-mode FOPA, which is similar to the model given in Sec. II, because both signal and idler beams can be in single temporal mode, respectively. The results illustrate the validity of the multi-mode theory. In Sec. V, we focus on analyzing the FOPA with a very broad gain bandwidth. In this situation, the JSF is non-factorable. The calculated results show that although the pulse pumped high gain FOPA generally has a noise figure greater than the 3 dB quantum limit, the intensity difference noise of the twin beams, weakly depends on the JSF, can be significantly reduced to lower than the SNL.
To closely resemble the real experimental situation, the dependence of the intensity difference noise of twin beams is numerically studied by considering the influences of experimental imperfections, including the quantum efficiencies of detectors, the collection efficiency of twin beams, and the excess noise of input signal. Finally, we conclude in Sec. \ref{Conclusion}.

\section{\label{SINGLE_MODE}Quantum description of a single mode FOPA\protect\\}
We start with the quantum description of single-mode linear FOPAs, whose pumps are single frequency lasers with negligible depletions. In this section, we will introduce the Bogoliubov transformation in the single frequency mode form and the analytical expressions of key parameters of the FOPA, including the photon number gain, the noise figure, and the intensity difference noise of the twin beams.

\begin{figure}[]
\includegraphics[width=0.65\textwidth]{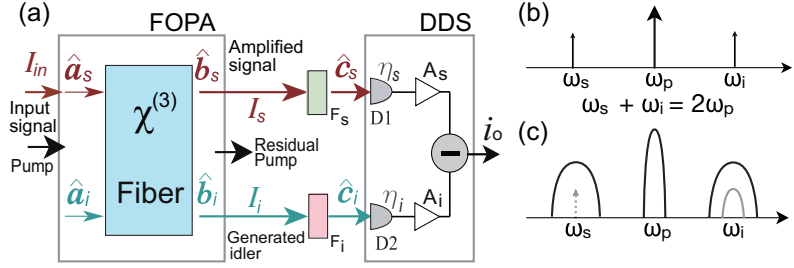}
 \caption{\label{FOPA}  (a) A typical diagram of the non-degenerate phase-insensitive FOPA. $F_s$ and $F_i$, filters; D1 and D2, detectors; $A_s$ and $A_i$, radio frequency amplifiers; DDS, differential detection system. (b) The sketch map of the frequency modes for the single frequency laser pumped FOPA.  Two pump photons of a single frequency laser uniquely defines the single frequency modes of a pair of signal and idler photons. (c) The sketch map of the frequency modes for the pulse pumped FOPA. A single frequency signal mode (dotted arrow) will be correlated to many idler frequency modes (enveloping curve in gray), and vice versa. Since the pump is in multi-frequency modes, the pulsed signal and idler beams are also in multi-frequency modes (enveloping curves in black).}
 \end{figure}

Figure \ref{FOPA}(a) is a typical diagram of the non-degenerate phase-insensitive FOPA, which is referred to as the FOPA hereinafter for brevity. When the wavelengths of the pump are properly set, the phase matching of FWM in fiber is satisfied. In the FWM process, two pump photons at frequencies $\omega_{p1}$ and $\omega_{p2}$ are simultaneously scattered into a pair of signal and idler photons at frequencies $\omega_{s}$ and $\omega_{i}$ via the $\chi^{(3)}$ nonlinearity, and the energy conservation relation $\omega_{p1}+\omega_{p2}=\omega_{s}+\omega_{i}$ is satisfied. Since each generated signal photon is always accompanied by the birth of an idler photon, the two form correlated twin beams.
A weak input is injected into the FOPA from either the signal or the idler field. For convenience, we refer the field with weak injection as the signal, while the other with vacuum input as idler. When the weak signal is amplified by the strong pump field via the FWM process, an idler beam is generated, and the frequency overlap between the amplified signal and generated idler fields is negligible for the non-degenerate FOPA.

When the FOPA is pumped with a single frequency CW laser, as illustrated in Fig. 1(b), the two pump photons have the same frequency mode ($\omega_{p1}= \omega_{p2}=\omega_{p}$), a pair of signal and idler photons in the frequency modes $\omega_{s}$ and $\omega_{i}$ are uniquely defined. In this case, the input-output relation of the single mode FOPA is given by the well-known Bogoliubov transformation:
  \begin{equation}
  \hat{b}_{s,i} = \mu \hat{a}_{s,i} +  \nu \hat a_{i,s}^{\dag}, ~~ \label{single_trans}
  \end{equation}
where  $a$ and $b$ refer the input and output modes, the subscript $s,i$ respectively denote the signal and idler fields, the coefficients $\mu$ and $\nu$ are determined by the parametric gain of FWM and satisfy the relation $|\mu|^2-|\nu|^2=1$.
For input signal in a coherent state $| \alpha \rangle_s^{(s)}$ with photon number much greater than 1, i.e.,
\begin{equation}
\label{input-sig}
  I_{in}=|\alpha|^2\gg1,
  \end{equation}
it is straightforward to deduce the expressions of all the parameters of the FOPA by calculating the quantum average of the corresponding operators over the state $| \Psi \rangle^{(s)}=| \alpha \rangle_s^{(s)} | 0 \rangle_i$, where $| 0 \rangle_i$ denotes the vacuum input of the idler field and the superscript ${(s)}$ denotes the single frequency mode case.

The amplified signal (generated idler) beam is then measured by the detector D1(2) after passing through the filter $F_{s(i)}$ to reject the strong pump (Fig. 1(a)). The average photon number $ I_{s(i)} $ and the photon number fluctuation $\Delta I_{s(i)}^2$ of the signal (idler) fields are converted to the DC and AC currents of D1 and D2, respectively. For identical D1 and D2, when the efficiencies of the detector D1(2) and filter $F_{s(i)}$ are ideal,
the average photon numbers of the amplified signal and generated idler beams can be written as
\begin{subequations}
\label{Sig_I}
\begin{equation}
I_s^{(s)}=\langle \hat b_{s}^{\dag}\hat b_{s} \rangle=|\mu|^2 | \alpha|^2+ |\nu|^2,
\end{equation}
\begin{equation}
I_i^{(s)}=\langle \hat b_{i}^{\dag}\hat b_{i} \rangle=|\nu|^2 | \alpha|^2+ |\nu|^2,
\end{equation}
\end{subequations}
where $\hat b_{s,(i)}^{\dagger} \hat b_{s,(i)}$ is the photon number operator of signal (idler) beam. In our theoretical model, we assume the condition $|\alpha|^2\gg1$ is always satisfied, so the contribution of spontaneous terms $|\nu|^2$ in the right hand side of Eqs. (\ref{Sig_I}a) and (\ref{Sig_I}b) are negligible.
According to the definition of the photon number gain---the ratio between the average photon numbers of amplified signal and input signal
\begin{equation}
\label{Def_g}
g =\frac{ I_{s} }{I_{in}},
\end{equation}
we have $g^{(s)}=|\mu|^2$. Moreover, according to the definition of the intensity noise of the amplified signal and generated idler fields, we have
\begin{subequations}
\label{Sig_Delta_I}
\begin{equation}
\Delta {I_{s}^{(s)}}^{2}=\langle (\hat b_{s}^{\dag}\hat b_{s})^2 \rangle-\langle \hat b_{s}^{\dag}\hat b_{s} \rangle^2 =(|\mu|^4+|\mu|^2|\nu|^2)|\alpha|^2
\end{equation}
\begin{equation}
\Delta {I_{i}^{(s)}}^{2}=\langle (\hat b_{i}^{\dag}\hat b_{i})^2 \rangle-\langle \hat b_{i}^{\dag}\hat b_{i} \rangle^2=(|\mu|^2|\nu|^2+|\nu|^4)|\alpha|^2.
\end{equation}
\end{subequations}
After using the SNL of signal (idler) beam $\Delta I_{cs(i)}^2=\langle \hat b_{s(i)}^{\dag}\hat b_{s(i)} \rangle$ to normalize the intensity noise $\Delta {I_{s(i)}^{(s)}}^{2}$, we obtain
\begin{equation}
\label{R_CW}
R_s^{(s)}={R_i^{(s)}}=2g^{(s)}-1,
\end{equation}
indicating the normalized intensity noise of the amplified signal beam $R_s^{(s)}$ is equal to that of the generated idler beam $R_i^{(s)}$. Notice that they are both above the SNL because of amplification.

As for the noise figure (NF) of the FOPA, according to the definition~\cite{Voss06}
\begin{equation}
\label{eq:NF1}
\emph{NF}=\frac{SNR_{in}}{SNR_{out}},
\end{equation}
with
\begin{equation}
\label{SNR-in}
SNR_{in}= \frac{I_{in}^2} { \Delta I_{in}^2}=|\alpha|^2
\end{equation}
and
\begin{equation}
\label{SNR-out}
SNR_{out}= \frac{I_{s}^2} { \Delta I_{s}^2}
\end{equation}
respectively denoting the signal to noise ratio of the input and amplified signal beams,
the NF of the single mode FOPA is given by
\begin{equation}
\label{NF_s_sm}
NF^{(s)}=1+\frac{|\nu|^2}{|\mu|^2}=\frac{2g^{(s)}-1}{g^{(s)}},
\end{equation}
which shows that the parametric amplification process adds the excess noise to the amplified signal beam and the 3 dB quantum limit of NF can be achieved in the high gain limit.

In the ideal case, the photocurrents of D1 and D2 are highly correlated and can be subtracted
from each other to yield quantum-noise reduction, i.e.,
$\frac{ \Delta ( \hat I_s- \hat I_i)^2}{\Delta I^2_{cs}+\Delta I^2_{ci} }<1 $.
In practice, the measured noise reduction is very sensitive to the collection efficiency of twin beams and the response function of differential detection system (DDS), consisting of D1 and D2 followed by the radio frequency (RF) amplifiers $A_s$ and $A_i$, and the subtractor (see Fig. 1(a)). For brevity, we assume throughout this paper that the transmission efficiencies of the filters $F_s$ and $F_i$ at their central wavelengths are perfect. In this situation, the collection efficiencies of the signal and idler beams at the single frequency $\omega_{s}$ and $\omega_{i}$ are perfect, so we only need to analyze the influence of DDS with non-ideal response.

The response function of the DDS is described by $Q_{1}(\Omega)$ and $Q_{2}(\Omega)$, which are determined by the quantum efficiencies and electrical circuit of D1 and D2, respectively. Here $\Omega$ denotes the radio frequency. The DC response of the DDS, $Q_{1(2)}(0)$, is associated with the quantum efficiency through the relation  $Q_{1(2)}(0)=\frac{e\eta_{s(i)}}{\hbar \omega} $, where $e$ is the electron charge, $\eta_{s(i)}$ is the quantum efficiency of D1 (D2) , and $\hbar$ is the reduced Planck constant; while the AC response, $Q_{1(2)}(\Omega)$ ($\Omega\neq 0$), is not only determined by the response of D1 (D2)  $D_{1(2)}{(\Omega)}$, but also determined by that of the RF amplifier $A_{s(i)}(\Omega)$. For convenience, the ratio $r$ between the AC response $Q_{1}(\Omega)$ and $Q_{2}(\Omega)$, defined as
\begin{eqnarray}
 r=\frac{\eta_s A_i(\Omega) D_2(\Omega)}{\eta_i A_s(\Omega) D_1(\Omega) }~~~(\Omega\neq0),
\end{eqnarray}
which can be adjusted by changing the electronic gain of the amplifier $A_{s}$ or $A_{i}$.

In practice, the quantum efficiencies of D1 and D2 are not perfect, i.e., $\eta_{s}< 1$ and $\eta_{i}< 1$. Modeling the non-ideal detectors as the beam splitters, which couple the vacuum mode $\hat v_s$ and $\hat v_i$ to the signal and idler fields, the field operator incident on D1 and D2 are given by
\begin{subequations}
\label{eq:sm_operator}
\begin{equation}
\hat c_s=\sqrt{\eta_s} \hat b_s+\sqrt{1-\eta_s}\hat v_s,
\end{equation}
\begin{equation}
\hat c_i=\sqrt{\eta_i} \hat b_i+\sqrt{1-\eta_i}\hat v_i.
\end{equation}
\end{subequations}
In this situation, the measured photon numbers of the amplified signal and idler beams in Eq. (3) are rewritten as
\begin{subequations}
\label{sm_Isi}
\begin{equation}
I_s^{(s)'}=\langle \hat c_{s}^{\dag}\hat c_{s} \rangle=\eta_s(|
\mu|^2 |\alpha|^2+ |\nu|^2)
\end{equation}
\begin{equation}
I_i^{(i)'}=\langle \hat c_{i}^{\dag}\hat c_{i} \rangle=\eta_i(|
\nu|^2 |\alpha|^2+ |\nu|^2),
\end{equation}
\end{subequations}
and the intensity noise of individual signal and idler beams in Eq. (5) is rewritten as
\begin{subequations}
\label{DeltaI_sm}
\begin{equation}
\Delta (I_s^{(s)'})^2=\langle (\hat c_{s}^{\dag}\hat c_{s})^2
\rangle-\langle (\hat c_{s}^{\dag}\hat c_{s}) \rangle^2=(2\eta_s^2|
\mu|^2 |\nu|^2 + \eta_s |\mu|^2)|\alpha|^2,
\end{equation}
\begin{equation}
\Delta (I_i^{(s)'})^2=\langle (\hat c_{i}^{\dag}\hat c_{i})^2
\rangle-\langle (\hat c_{i}^{\dag}\hat c_{i}) \rangle^2=(2\eta_i^2|
\nu|^4+ \eta_i |\nu|^2)|\alpha|^2.
\end{equation}
\end{subequations}

For the operator of the measured intensity difference of the twin beams
\begin{equation}
\hat{I}_t^{(s)'}=\hat c_{s}^{\dag}\hat c_{s}-r\hat c_{i}^{\dag}
\hat c_{i},
\end{equation}
the normalized intensity difference noise of the twin beams is expressed as
\begin{equation}
  \label{eq:electrical_gain}
R_t'^{(s)}=\frac{\Delta (I_t^{(s)'})^2}{\Delta (I_{cs}
^{(s)'})^2+r^2 \Delta (I_{ci}^{(s)'})^2}=\frac{\Delta
(I_s^{(s)'})^2+r^2\Delta (I_i^{(s)'})^2-2r H_{si}}{\Delta (I_{cs}
^{(s)'})^2+r^2 \Delta (I_{ci}^{(s)'})^2},
  \end{equation}
where $\Delta (I_{cs(i)}^{(s)’})^2=I_{s(i)}^{(s)’}$ is the corresponding SNL of
the detected signal (idler) beam and $H_{si}=2 \eta_s \eta_i |\mu|^2 |\nu|^2 |\alpha|^2$ is the quantum correlation term.
Substituting Eqs. (\ref{sm_Isi})-(\ref{DeltaI_sm}) into Eq. (\ref{eq:electrical_gain}), we obtain
\begin{equation}
\label{eq:R_sm_eta}
R_{t}'^{(s)}(\eta_s,\eta_i,r)=\frac{ \eta_i \big( 2\eta_i|\nu|^4+|\nu|^2\big)r^2 -4\eta_s \eta_i |\mu|^2|\nu|^2r + 2\eta_s^2 |\mu|^2|\nu|^2 + \eta_s |\mu|^2} { \eta_s |\mu|^2 + \eta_i |\nu|^2 r^2 }.
\end{equation}

To illustrate the factors influencing the value of $R_{t}'^{(s)}(\eta_s,\eta_i,r)$, we first plot $R_{t}'^{(s)}$ as a function of the gain $g^{(s)}$ by assuming the ratio $r$ in Eq. (11) is equal to 1 and by varying the efficiencies $\eta_s$ and $\eta_i$, as shown in Fig. 2(a). When the efficiencies of both D1 and D2 are $75\%$ or $85\%$, $R_t'^{(s)}$ decreases with the increase of $g^{(s)}$; for a given $g^{(s)}$, $R_t'^{(s)}$ obtained for $\eta_{s}=\eta_{i}=85\%$ is lower than that for $\eta_{s}=\eta_{i}=75\%$. When efficiencies of D1 and D2 are $\eta_s=75\%$ and $\eta_i=85\%$, respectively, $R_t'^{(s)}$ is higher than that for $\eta_s=\eta_i=75\%$ in the high gain regime, however, $R_t'^{(s)}$ can be slightly lower than that for $\eta_s=\eta_i=85\%$ when the gain is within the regime of $3<g^{(s)}<10$. Therefore, for the case of $\eta_s=\eta_i$, $R_t'^{(s)}$ decreases with $g^{(s)}$ and increases with the decrease of $\eta_{s(i)}$; while for the case of $\eta_s\neq\eta_i$, $R_t'^{(s)}$ does not always decrease with $g^{(s)}$: after obtaining the minimum $R_t'^{(s)}$ at a certain gain, $R_t'^{(s)}$ will start to increase with $g^{(s)}$.

\begin{figure}[]
\includegraphics[width=0.55\textwidth]{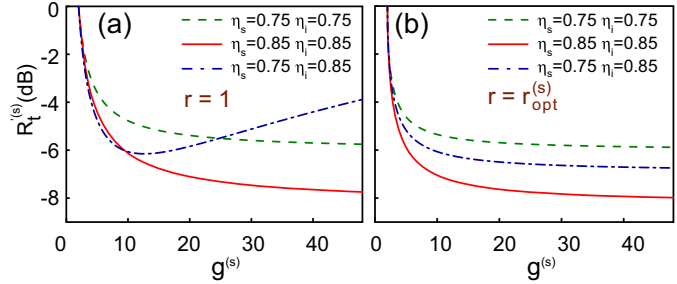}
 \caption{\label{Rt-single-mode} (a) The noise reduction $R_t'^{(s)}$ as a function of the gain $g^{(s)}$ under different quantum efficiencies for the cases of (a) $r=1$ and (b) $r=r_{opt}^{(s)}$. }
 \end{figure}

Eq. (\ref{eq:R_sm_eta}) shows that for the given values of $\eta_s$, $\eta_i$ and $g^{(s)} $, $R_t'^{(s)}$ can be minimized  when the ratio $r$ takes the optimized value
\begin{equation}
\label{roptsm}
r_{opt}^{(s)}=\frac{1}{2}(\frac{\eta_s}{\eta_i}-1)+ \frac{1}{2\eta_i} \sqrt{\frac{g^{(s)}(\eta_s+\eta_i)^2+(\eta_s-\eta_i)^2}{g^{(s)}-1}},
\end{equation}
We then plot $R_t'^{(s)}$ as a function of $g^{(s)}$ by varying the efficiencies $\eta_s$ and $\eta_i$ for the case of $r=r_{opt}^{(s)}$. As shown in Fig. 2(b), no matter $\eta_s$ and $\eta_i$ are equal or not, $R_t'^{(s)}$ always decreases with the increase of $g^{(s)}$. Moreover, for a fixed value of $g^{(s)}$, our calculation indicates that $R_t'^{(s)}$ decreases with the increase of $\eta_{s}$ and $\eta_{i}$.

We note that it is impossible to obtain the ideal noise reduction $R_{t}'^{(s)}(\eta_s,\eta_i,r)=0$ for the input signal with photon number $|\alpha|^2\neq0$, even if the detectors D1 and D2 are perfect. To clearly illustrate this point, let's look at the expressions of $R_{t}'^{(s)}$ when the condition $\eta_s=\eta_i=1$ is fulfilled.
For the case of $r=r_{opt}^{(s)} $, Eq. (\ref{eq:R_sm_eta}) is rewritten as
\begin{equation}
\label{R_sm_r}
R_{t}'^{(s)}(1,1;r_{opt})=\frac{1}{\big(|\mu|+|\nu|\big)^2},
\end{equation}
while for the case of $r=1$, Eq. (\ref{eq:R_sm_eta}) is rewritten as
\begin{equation}
\label{R_t_sm}
  R_t'^{(s)}(1,1;1)=\frac{1}{|\mu|^2+|\nu|^2}.
\end{equation}
Both Eqs. (\ref{R_sm_r}) and (\ref{R_t_sm}) show the intensity difference noise of twin beams is always less than the SNL (1 or $0$ dB), however, the condition $R_t'^{(s)} > 0$ is always satisfied. Moreover, a comparison between Eqs. (\ref{R_sm_r}) and (\ref{R_t_sm}) indicates that in the high gain limit, the noise reduction can be improved up to 3 dB by optimizing the ratio $r$.

\section{\label{pulsed FOPA} Quantum description of a pulse pumped FOPA\protect\\}
When the FOPA is pumped by a strong pulsed field having multiple frequency components, such as a mode-locked laser with a pulse width of a few pico-seconds, a single frequency signal mode is correlated to many idler frequency modes, and vice versa (see the dotted arrow and the corresponding enveloping curve in Fig.1 (c)). In this situation, both amplified signal and generated idler beams are pulsed fields with broad bandwidth, as illustrated by the spectrally broadened curves in Fig. 1(c). Therefore, it is necessary to deal with the multi-frequency modes. In this section, we first extend the single mode Bogoliubov transformation to the multi-frequency domain. Then, according to the Hamiltonian of FWM in fiber, we deduce the four transformation functions of the multi-mode Bogoliubov transformation, from which the general expressions of the key parameters of a pulse pumped high gain FOPA can be derived.

\subsection{\label{multi_mode_general}Multi-mode model of a pulse pumped FOPA}

An optical field propagating along the fiber (denoted as the z direction) can be quantized by using one dimensional approximation if the polarization mode is well defined. Therefore, the pulsed field operator in optical fiber is written as \cite{Ou1995}
\begin{equation}
  \hat{E}(t)=\hat{E}^{(+)}(t)+\hat{E}^{(-)}(t),
  \end{equation}
where the positive frequency field operator $\hat{E}^{(+)}(t)$ and negative frequency operator $\hat{E}^{(-)}(t)$ satisfy the relation $\hat{E}^{(+)}(t)=[\hat{E}^{(-)}(t)]^\dagger$,
and $\hat{E}^{(+)}(t)$ is given by
  \begin{equation}
  \hat{E}^{(+)}(t)=\frac{1}{\sqrt{2 \pi}} \int d\omega \hat{a}(\omega)\textit{e}^{ \textit{i}(k \cdot z-\omega t)} = \frac{1}{\sqrt{2 \pi}} \int d\omega \hat{a}(\omega)\textit{e}^{ -\textit{i}\omega t'}.
  \end{equation}
Here the annihilation operator at frequency $\omega$, $\hat{a}(\omega)$, satisfies the commutation relation $[\hat{a}(\omega),\hat{a}^{\dag}(\omega')]=\delta(\omega-\omega')$. Note that $t'\equiv t- kz/\omega$, so the distance in propagation is equivalent to a delay in time.

For the pulse pumped FOPA, the strong pump pulses remain classical, but the signal and idler fields are quantized.
Extending Bogoliubov transformation in Eq. (\ref{single_trans}) to the multi-frequency domain and using Eq. (22), we write the input-output relation as~\cite{Wasi06}:
\begin{subequations}
\label{eq:multimode_trans}
\begin{equation}
\hat{b}_s(\omega_s)= \hat{U}^{\dag} \hat{a}_s(\omega_s') \hat{U}=\int_S h_{1s}(\omega_s,\omega_s')\hat{a}_s(\omega_s') d\omega_s'+ \int_I h_{2s}(\omega_s,\omega_i')\hat{a}_i^{\dag}(\omega_i') d\omega_i'
\end{equation}
\begin{equation}
\hat{b}_i(\omega_i)=\hat{U}^{\dag} \hat{a}_i(\omega_i') \hat{U}=\int_I h_{1i}(\omega_i,\omega_i')\hat{a}_i(\omega_i') d\omega_i'+ \int_S h_{2i}(\omega_i,\omega_s')\hat{a}_s^{\dag}(\omega_s') d\omega_s' ,
\end{equation}
\end{subequations}
where the unitary evolution operator $\hat{U}$ is determined by the Hamiltonian of the FOPA(see later for the explicit form), the footnotes $S,I$ denote the frequency ranges of the signal and idler fields, and the four Green functions $h_{nj}(\omega,\omega')\ (n=1,2\ j=s,i)$ are referred to as the transformation functions. To ensure the satisfaction of the commutation relations of the operators $\hat a_{s(i)}(\omega_{s(i)}')$ and $\hat b_{s(i)}(\omega_{s(i)})$, the transformation functions should be constrained by the relations:
  \begin{subequations}
  \label{eq:FWM_OPA_Unitarity_Conditions}
  \begin{eqnarray}
  &&\int_S h_{1s}(\omega_s, \omega_s') h_{2i}(\omega_i, \omega_s') d \omega_s' - \int_I h_{2s}(\omega_s,  \omega_i') h_{1i}(\omega_i, \omega_i') d\omega_i' =0
  \\
 &&\int_S h_{1s}(\omega_{s1}, \omega_s') h_{1s}^*(\omega_{s2}, \omega_s') d \omega_s' - \int_I h_{2s}(\omega_{s1},  \omega_i') h_{2s}^*(\omega_{s2}, \omega_i') d\omega_i' =\delta(\omega_{s1}-\omega_{s2})
 \\
  &&\int_I h_{1i}(\omega_{i1}, \omega_i') h_{1i}^*(\omega_{i2}, \omega_i') d \omega_i' - \int_S h_{2i}(\omega_{i1},  \omega_s') h_{2i}^*(\omega_{i2}, \omega_s') d\omega_s' =\delta(\omega_{i1}-\omega_{i2}).
  \end{eqnarray}
  \end{subequations}

The weak input signal pulses, synchronized with the pump pulses, is ideally in a multi-mode coherent state
  \begin{equation}
  \label{eq:Coh_State}
  | \alpha \rangle _s = | \{\alpha(\omega_s)\} \rangle = \exp\Big\{ \int _0^{\infty }[ \alpha\cdot s(\omega_s)\hat{a}^{\dag}-\emph{h.c.} ]d\omega_s \Big\}  | 0 \rangle,
  \end{equation}
where $\alpha$ is the complex amplitude, and the frequency distribution function $s(\omega_s)$ satisfies the normalization condition $\int_0^{\infty}|s(\omega_s)|^2 d\omega_s=1$ so that the photon number of the input signal is $I_{in}=|\alpha|^2$, which is the same as Eq. (\ref{input-sig}).
In general, the bandwidth of the input signal is much smaller than that of the central frequency. So the quasi-monochromatic approximation applies, and the frequency integral range from $0$ to $\infty$ can be treated as from $-\infty$ to $\infty$. For the sake of brevity, all the integral ranges from $-\infty$ to $\infty$ will be omitted hereinafter.

In contrast to the single mode FOPA, wherein the filters $F_{s}$ and $F_{i}$ with ideal transmission efficiency at the central frequencies result in perfect collection efficiency of twin beams, the collection efficiency of the broad band multi-mode twin beams is associated with the spectra of $F_s$ and $F_i$. In this case, the field operators incident on the detectors D1 and D2 are given by
\begin{subequations}
\label{eq:vacuum_mixing}
\begin{equation}
\hat c_s (\omega_s ) = \sqrt {\eta _s } f_s (\omega_s )\hat b_s (\omega_s ) + i\sqrt {1 - \eta _s f_s^2 (\omega_s )} \hat v_s (\omega_s )
\end{equation}
\begin{equation}
\hat c_i (\omega_i ) = \sqrt {\eta _i } f_i (\omega_i )\hat b_i (\omega_i ) + i\sqrt {1 - \eta _i f_i^2 (\omega_i )} \hat v_i (\omega_i ),
\end{equation}
\end{subequations}
where the complex function $f_j(\omega_j)$ ($j=s,i$) describes the spectrum of the filter $F_j$. Eq.  (\ref{eq:vacuum_mixing}) implies that the measured quantum noise property of the pulse pumped FOPA is not only influenced by the quantum efficiencies of D1 and D2, but also influenced by the spectral functions $f_s(\omega_s)$ and $f_i(\omega_i)$. For the ideal detectors, we should have $\eta_s=\eta_i=1$, while for the perfect collection of the twin beams, we should have $|f_s(\omega_s)|=|f_i(\omega_i)|=1$.

Similar to the single mode FOPA, the key parameters of the pulse pumped FOPA can be obtained by calculating the quantum average of the corresponding operators over the state $| \Psi \rangle=| \alpha \rangle_s | 0 \rangle_i$. In the next subsection, we will derive the analytical forms of $h_{nj}(\omega,\omega')\ (n=1,2\ j=s,i)$, which are the key for deriving the formulas of the operators of twin beams (see Eq. (\ref{eq:multimode_trans})).

\subsection{\label{Multi_mode_filber_OPA} Derivation of the transformation function}

The unitary operator in Eq. (\ref{eq:multimode_trans}) is determined by the Hamiltonian of parametric process. So let's begin with
the Hamiltonian of the co-polarized pulse pumped FWM in single-mode optical fibers~\cite{Stolen82,Chen05}
\begin{equation}
  \label{eq:Hamiltonian_0}
  \hat{H}(t)= C_1 \chi^{(3)} \int dV[E_{p1}(t) E_{p2}(t) \hat{E}_{s}^{(-)}(t) \hat{E}_{i}^{(-)}(t) + \emph{h.c.}],
  \end{equation}
where $C_1$ is a constant determined by experimental details and the units of quantized optical fields, $\chi^{(3)}$ is the 3rd-order real nonlinearity, $E_{p1}$ and $E_{p2}$ are the classical pump fields,
\begin{subequations}
  \label{eq:Ea}
  \begin{equation}
  \hat{E}_{s}^{(-)}(t)=\frac{1}{\sqrt{2\pi}} \int d\omega_s' \hat{a}^{\dag}_s(\omega_s')\textit{e}^{ -\textit{i}(k_s z-\omega_s' t)}
  \end{equation}
and
  \begin{equation}
  \hat{E}_{i}^{(-)}(t)=\frac{1}{\sqrt{2\pi}} \int d\omega_i' \hat{a}^{\dag}_i(\omega_i')\textit{e}^{ -\textit{i}(k_i z-\omega_i' t)}
  \end{equation}
  \end{subequations}
are the quantized negative-frequency field operator of the signal and idler beams, respectively.
The Gaussian shaped strong pump pulses propagating along the
fiber can be written as \cite{Chen05,Yang2011}
\begin{equation}
\label{eq:pump}
E_{pn}(t) =E_{0} \emph{e}^{-i \gamma P_p z} \int \emph{e}^{-(\omega_{pn}-\omega_{p0})^2/2\sigma_p^2} \emph{e}^{i(k_{pn} z-\omega_{pn} t)} d\omega_{pn}~~ (n=1,2),
\end{equation}
where $\sigma_p$, $\omega_{p0}$ and $k_{p}$ are the bandwidth, central frequency, and wave vector of the pump, respectively, and $E_0$ is related to the peak power through the relation $P_p=2 \pi \sigma_p^2 E_0^2 $. The nonlinear coefficient $\gamma$ is expressed as $\gamma=\frac{3\omega_{p0}\chi^{(3)}}{8 c A_{eff}}$, where $A_{eff}$ denotes the effective mode area and $c$ is the speed of light in vacuum. After substituting Eqs. (\ref{eq:Ea}) and (\ref{eq:pump}) into Eq. (\ref{eq:Hamiltonian_0}) and changing $dV$ in Eq. (\ref{eq:Hamiltonian_0}) to $dV=A_{eff} dz$, we carry out the integral over the whole fiber length $L$ (from -L/2 to L/2) and arrive at the Hamiltonian in the time dependent form
\begin{eqnarray}
\label{H}
\hat{H}(t)=&&\frac{2 C_1 \gamma P_p L c A_{eff}^2}{3 \omega_{p0} \pi^2 \sigma_p^2}\int d\omega_{p1} d\omega_{p2} d\omega_{s}' d\omega_{i}'\hat{a}^{\dag}_s(\omega_s') \hat{a}^{\dag}_i(\omega_i')\nonumber\\
&&sinc\big(\frac{\Delta k L}{2} \big) \exp \big\{ -\frac{ (\omega_{p1}-\omega_{p0})^2+(\omega_{p2}-\omega_{p0})^2 } {2\sigma_p^2} \big\} \emph{e}^{-\emph{i}(\omega_{p1}+\omega_{p2}-\omega_{s}'-\omega_{i}')t}+\emph{h.c.}
\end{eqnarray}
where $\Delta k=k_s+k_i-2k_p+2\gamma P_p$ is the phase mismatching term, and the term $2\gamma P_p$ is originated from the self-phase modulation of pump.

To obtain the formula of the unitary evolution operator $\hat{U}=\exp{\{\frac{\int \hat{H}(t) dt}{i\hbar}\}} $, we need to carry out the integration over the time and over all the possible combinations of $\omega_{p1}$ and $\omega_{p2}$ within the pump bandwidth. Since the time integral gives rise to the $\delta$ function in $\Delta\omega=\omega_{p1}+\omega_{p2}-\omega_{s}'-\omega_{i}'$ to
guarantee the energy conservation at single-photon level, we arrive at
  \begin{equation}
\label{eq:nondegererate_evolution_operator}
  \hat{U}=\exp \Big\{ G[\iint \psi(\omega_s',\omega_i') \hat{a}_s^\dag(\omega_s') \hat{a}_i^\dag(\omega_i') d\omega_s' d\omega_i' - \emph{h.c.}]\Big\},
  \end{equation}
where the so called two-photon joint spectral function (JSF) of FWM has the form of~\cite{Palmett07}
\begin{equation}
\label{JSF_def}
\psi(\omega_s' ,\omega_i')=\frac{C}{2\sqrt{\pi}\sigma_p}\exp\big \{ \frac{-(\omega_s'+\omega_i'-2\omega_{p0})^2}{4\sigma_p^2}\big \}  \emph{sinc}(\frac{\Delta k L}{2}),
  \end{equation}
which is determined by the pump envelop function $\exp\big \{ \frac{-(\omega_s'+\omega_i'-2\omega_{p0})^2}{4\sigma_p^2} \big \}$ and phase matching function $\emph{sinc}(\frac{\Delta k L}{2})$, and is referred to as the probability amplitude of simultaneously finding a pair of signal and idler photons within the frequency range of $\omega_s' \rightarrow \omega_s'+d\omega_s'$ and $\omega_i' \rightarrow \omega_i'+d\omega_i'$, respectively. The coefficient $G= \frac{-8 \emph{i} C_1 \gamma P_p L c A_{eff}^2}{3 \hbar \omega_{p0} C} $ in Eq. (\ref{eq:nondegererate_evolution_operator}) determines the gain of FWM, and
the coefficient $C$ in Eq. (\ref{JSF_def}) is a constant used to ensure the satisfaction of the normalization condition
$\iint |\psi(\omega_s',\omega_i')|^2 d\omega_s' d\omega_i'=1$. According to Eq. (\ref{JSF_def}), the JSF is generally asymmetry since $\Delta k$ is frequency dependent. Therefore, $\psi(\omega_1,\omega_2) =\psi(\omega_2,\omega_1)$ is generally not satisfied.

To obtain the formula of the transformation functions, we rewrite the input-output relation by substituting Eqs. (\ref{eq:nondegererate_evolution_operator}) into Eqs.(\ref{eq:multimode_trans}) and using Baker-Hausdorff lemma:
\begin{eqnarray}
   \hat{b}_j(\omega_j)&&=\hat{U}^{\dag} \hat{a}_{j}(\omega_j') \hat{U}=\hat{a}_j(\omega_j') + [\hat{a}_{j}(\omega_j'),\hat{B}]\nonumber\\
 &&\ \ +\frac{1}{2!}[\hat{a}_j(\omega_j'),[\hat{a}_j(\omega_j'),\hat{B}]]+\frac{1}{3!}[\hat{a}_j(\omega_j'),[\hat{a}_j(\omega_j'),[\hat{a}_j(\omega_j'),\hat{B}]]]+...\ ,~~ \label{LieSeries}
   \end{eqnarray}
  where the operator $\hat B$ is expressed as
   \begin{equation}
   \hat{B}=G\iint [\psi(\omega_s',\omega_i') \hat{a}_s^\dag(\omega_s') \hat{a}_i^\dag(\omega_i') d\omega_s' d\omega_i' - \emph{h.c.}].
   \end{equation}
Then the transformation functions can be written in the form of an infinite series
\begin{subequations}
\label{eq:Trans_Intergral}
\begin{eqnarray}
h_{1s}(\omega_s,\omega_s')= &&\delta(\omega_s-\omega_s')+\sum_{n=1}^{\infty} \frac{G^{2n}}{(2n)!}\iint\cdot\cdot\cdot\int d\omega_1d\omega_2\cdot\cdot\cdot d\omega_{2n-1}\nonumber\\
 \{ &&[\psi(\omega_s,\omega_1)\psi(\omega_2,\omega_3)\psi(\omega_4,\omega_5) \cdot\cdot\cdot \psi(\omega_{2n-2},\omega_{2n-1})]\nonumber\\ &&[\psi^*(\omega_2,\omega_1)\psi^*(\omega_4,\omega_3)\psi^*(\omega_6,\omega_5) \cdot\cdot\cdot \psi^*(\omega_s',\omega_{2n-1})] \}
\end{eqnarray}
\begin{eqnarray}
h_{2s}(\omega_s,\omega_i')=&&G\psi(\omega_s,\omega_i')+\sum_{n=1}^{\infty} \frac{G^{2n+1}}{(2n+1)!}\iint\cdot\cdot\cdot\int d\omega_1d\omega_2\cdot\cdot\cdot d\omega_{2n}\nonumber\\
 \{&& [\psi^*(\omega_2,\omega_1)\psi^*(\omega_4,\omega_3) \cdot\cdot\cdot \psi^*(\omega_{2n},\omega_{2n-1})]\nonumber\\&&[\psi(\omega_s,\omega_1)\psi(\omega_2,\omega_3)\psi(\omega_4,\omega_5) \cdot\cdot\cdot \psi(\omega_{2n},\omega_i')] \}
\end{eqnarray}
\begin{eqnarray}
h_{1i}(\omega_i,\omega_i')= &&\delta(\omega_i-\omega_i')+\sum_{n=1}^{\infty} \frac{G^{2n}}{(2n)!}\iint\cdot\cdot\cdot\int d\omega_1d\omega_2\cdot\cdot\cdot d\omega_{2n-1}\nonumber\\
&& \{ [\psi(\omega_1,\omega_i)\psi(\omega_3,\omega_2)\psi(\omega_5,\omega_4) \cdot\cdot\cdot \psi(\omega_{2n-1},\omega_{2n-2})]\nonumber\\ &&[\psi^*(\omega_1,\omega_2)\psi^*(\omega_3,\omega_4)\psi^*(\omega_5,\omega_6) \cdot\cdot\cdot \psi^*(\omega_{2n-1},\omega_i')] \}
\end{eqnarray}
\begin{eqnarray}
h_{2i}(\omega_i,\omega_s')= &&G\psi(\omega_s',\omega_i)+\sum_{n=1}^{\infty} \frac{G^{2n+1}}{(2n+1)!}\iint\cdot\cdot\cdot\int d\omega_1d\omega_2\cdot\cdot\cdot d\omega_{2n} \nonumber \\  \{ &&[\psi^*(\omega_1,\omega_2)\psi^*(\omega_3,\omega_4) \cdot\cdot\cdot \psi^*(\omega_{2n-1},\omega_{2n})]\nonumber \\
&&[\psi(\omega_1,\omega_i)\psi(\omega_3,\omega_2)\psi(\omega_5,\omega_4) \cdot\cdot\cdot \psi(\omega_s',\omega_{2n})] \}.
\end{eqnarray}
\end{subequations}
Eq. (\ref{eq:Trans_Intergral}) clearly shows that the transformation functions satisfy the relations
  \begin{subequations}
  \label{eq:signal_idler_Correlation}
  \begin{equation}
  h_{1s}(\omega_s,\omega_s')=h_{1s}^*(\omega_s',\omega_s)
  \end{equation}
  \begin{equation}
  h_{1i}(\omega_i,\omega_i')=h_{1i}^*(\omega_i',\omega_i)
  \end{equation}
  \begin{equation}
  h_{2s}(\omega_s,\omega_i')=h_{2i}(\omega_i',\omega_s),
  \end{equation}
  \end{subequations}
indicating that there is a correlation between the signal and idler beams.

It is worth noting that Bogoliubov transformation in multi-frequency modes (Eq (\ref{eq:multimode_trans})) with the transformation function shown in Eq. (\ref{eq:Trans_Intergral}) is also suitable for describing the single mode FOPA. For example, if the JSF in Eq. (\ref{JSF_def}) takes the limit of a single frequency laser pumped case, i.e., $\psi(\omega_s',\omega_i')=\delta ( (\omega_s'-\omega_{s0})+(\omega_i'-\omega_{i0}))$, with $\omega_{s0}$ and $\omega_{i0}$ denoting the single frequencies of the signal and idler beams, the input-output relation of Eq (\ref{eq:multimode_trans}) would transform into $ \hat{b}_{s(i)} = \mu \hat{a}_{s(i)} +  \nu \hat a_{i(s)}^{\dag}$ with $ \mu =\cosh G$ and $\nu =\sinh G$, which is exactly the same as Eq. (\ref{single_trans}). Hence, Eqs. (\ref{eq:multimode_trans}) and (\ref{eq:Trans_Intergral}) can be viewed as a general description of the input-output relation of an FOPA.

\subsection{\label{Gain} Key parameters of the pulse pumped FOPA}

For the measurements of the signal and idler fields, in addition to taking the quantum efficiencies, $\eta_s$ and $\eta_i$, and the spectrum of filter $F_{s(i)}$ into account, we need to consider the response times of the detectors D1 and D2 as well. When the response time of each detector is much longer than the pump pulse duration, but is much shorter than the period of two adjacent pump pulses, the average photon number of the amplified signal (generated idler) beam per pulse can be expressed as
\begin{equation}
\label{I_j}
I_{s(i)}=\langle \hat{I}_{s(i)} \rangle =\int \langle \Psi | \hat{I}_{s(i)} (t) | \Psi \rangle dt,
\end{equation}
where
\begin{equation}
\label{number-operator}
\hat I_{j}(t)=\hat{E}_j^{(-)}(t)\hat{E}_j^{(+)}(t)
\end{equation}
 with
\begin{equation}
\label{field-operator}
 \hat{E}_j^{(+)}(t)=\frac{1}{\sqrt{2\pi}} \int  \hat{c}_j(\omega_j)\textit{e}^{ \textit{i}(k_j z-\omega_j t)} d\omega_j ~~(j=s,i)
\end{equation}
is the photon number operator of the pulsed field. The time integral ranges in Eqs. (\ref{I_j}) and (\ref{field-operator}) are omitted, because they can be treated as from $-\infty$ to $\infty$.

With the filter operator incident on D1 and D2 (Eq. (\ref{eq:vacuum_mixing})),  we substitute Eqs. (\ref{eq:multimode_trans}), (\ref{eq:Coh_State}), (\ref{field-operator}) into Eq. (\ref{I_j}), and obtain the expressions of the detected photon number
\begin{subequations}
  \label{eq:signal_idler_Average Photon Number}
  \begin{eqnarray}
  \label{eq:Is}
  I_s=&& |\alpha|^2 \eta_s \int d\omega_s |f_s(\omega_s )|^2 S_s(\omega_s)\nonumber\\
 + && \eta_s  \int d\omega_s \int  d\omega_i' |f_s (\omega_s)|^2 |h_{2s} (\omega_s,\omega_i' )|^2
  \end{eqnarray}
  \begin{eqnarray}
 I_i= &&  |\alpha|^2 \eta_i \int d\omega_i |f_i(\omega_i)|^2 S_i(\omega_i)\nonumber\\
 +&& \eta_i \int d\omega_i \int d\omega_s' |f_i (\omega_i )|^2 |h_{2i} (\omega_i ,\omega_s' )|^2,
  \end{eqnarray}
  \end{subequations}
where
\begin{subequations}
\label{spectra}
\begin{equation}
S_s(\omega_s)=\iint h_{1s}^* (\omega_s ,\omega _{s1}' )h_{1s} (\omega_s ,\omega _{s2}' )s^* (\omega _{s1}' )s (\omega _{s2}' ) d\omega _{s1}' d\omega _{s2}'
\end{equation}
and
\begin{equation}
S_i(\omega_i)=\iint  h_{2i}^* (\omega_i ,\omega _{s1}' )h_{2i} (\omega_i ,\omega _{s2}' )s^* (\omega _{s2}' )s (\omega _{s1}' )d\omega_{s1}' d\omega _{s2}'
\end{equation}
\end{subequations}
are the power spectra of the signal and idler beams.
The first terms in the right hand side of Eqs. (\ref{eq:signal_idler_Average Photon Number}a) and (\ref{eq:signal_idler_Average Photon Number}b) are originated from the stimulated emission of input signal, while the second terms are from the spontaneous emission. Since we have assumed that the photon number of the weak input signal per pulse is much greater than 1,
the spontaneous emission terms are negligible.

Using Eq. (\ref{eq:Is}) and the definition of photon number gain in Eq. (\ref{Def_g}), we find the expression of the photon number gain of the pulse pumped FOPA is
\begin{equation}
\label{gDef}
g=\int S_s(\omega_s) d\omega_s
\end{equation}
for the case of $\eta_s=1$ and $|f_s(\omega_s)|=1$.

To compute the intensity noise of the amplified signal (generated idler) beam
\begin{equation}
\label{eq:fluctuation_general}
\Delta I_{s(i)} ^2 =\langle \hat{I}_{s(i)}\hat{I}_{s(i)} \rangle-I_{s(i)}^2= \iint \langle \Psi | \hat{I}_{s(i)}(t_1)\hat{I}_{s(i)}(t_2) |\Psi \rangle   dt_1 dt_2- I_{s(i)}^2,
\end{equation}
we substitute Eqs. (\ref{eq:multimode_trans}),  (\ref{eq:Coh_State}), (\ref{eq:vacuum_mixing}) and  (\ref{I_j})-(\ref{field-operator}) into Eq. (\ref{eq:fluctuation_general}), and arrive at the detailed expression
\begin{subequations}
\label{eq:Photon_number_fluctuation_FWM_OPA}
\begin{equation}
\Delta I_s^2=\big( H_{s1}+H_{s2}+H_{vs} \big) |\alpha|^2
\end{equation}
\begin{equation}
\Delta I_i^2= \big( H_{i1}+H_{i2}+H_{vi} \big) |\alpha|^2,
\end{equation}
\end{subequations}
with
\begin{subequations}
\label{eq45}
\begin{eqnarray}
\label{eq:H_s1}
H_{s1} =&&\eta_s^2 \iint\ldots\int d\omega_{s1} d\omega_{s2} d\omega_{s1}'d\omega_{s2}'d\omega_{s3}'\nonumber\\
&&\{|f_s(\omega_{s1})|^2|f_s(\omega_{s2})|^2 h^*_{1s}(\omega_{s1},\omega_{s1}') h_{1s}(\omega_{s2},\omega_{s1}')\nonumber\\ &&h^*_{1s}(\omega_{s1},\omega_{s2}')s^*(\omega_{s2}')h_{1s}(\omega_{s2},\omega_{s3}')s(\omega_{s3}')\}
\end{eqnarray}
\begin{eqnarray}
\label{eq:H_s2}
H_{s2} =&& \eta_s^2 \iint\ldots\int d\omega_{s1} d\omega_{s2} d\omega_{i1}'d\omega_{s1}'d\omega_{s2}'\nonumber\\
&&\{|f_s(\omega_{s1})|^2|f_s(\omega_{s2})|^2 h^*_{2s}(\omega_{s1},\omega_{i1}') h_{2s}(\omega_{s2},\omega_{i1}')\nonumber\\
&&h^*_{1s}(\omega_{s2},\omega_{s1}')s^*(\omega_{s1}')h_{1s}(\omega_{s1},\omega_{s2}')s(\omega_{s2}')\}
\end{eqnarray}
\begin{eqnarray}
\label{eq:H_i1}
H_{i1} =&& \eta_i^2 \iint\ldots\int d\omega_{i1} d\omega_{i2} d\omega_{s1}'d\omega_{s2}'d\omega_{s3}'\nonumber\\
&&\{|f_i(\omega_{i1})|^2|f_i(\omega_{i2})|^2 h^*_{2i}(\omega_{i1},\omega_{s1}') h_{2i}(\omega_{i2},\omega_{s1}')\nonumber\\ &&h_{2i}(\omega_{i1},\omega_{s2}')s^*(\omega_{s2}')h^*_{2i}(\omega_{i2},\omega_{s3}')s(\omega_{s3}')\}
\end{eqnarray}
\begin{eqnarray}
\label{eq:H_i2}
H_{i2} =&& \eta_i^2 \iint\ldots\int d\omega_{i1} d\omega_{i2} d\omega_{i1}'d\omega_{s1}'d\omega_{s2}'\nonumber\\
&&\{|f_i(\omega_{i1})|^2|f_i(\omega_{i2})|^2 h_{1i}(\omega_{i1},\omega_{i1}') h^*_{1i}(\omega_{i2},\omega_{i1}')\nonumber\\ &&h_{2i}(\omega_{i2},\omega_{s1}')f^*(\omega_{s1}')h^*_{2i}(\omega_{i1},\omega_{s2}')f(\omega_{s2}')\}
\end{eqnarray}
\end{subequations}
and
\begin{subequations}
\label{eq46}
\begin{eqnarray}
\label{eq:H_vac_s}
H_{vs} = && \eta_s \int d\omega_s (1-\eta_s |f_s(\omega_s)|^2)\nonumber\\
\ \ \ \ &&\iint d\omega_{s1}' d\omega_{s2}' [h_{1s}^* (\omega_s ,\omega_{s1}' )h_{1s} (\omega_s ,\omega_{s2}' )s^* (\omega _{s1}' )s (\omega _{s2}' )]
\end{eqnarray}
\begin{eqnarray}
\label{eq:H_vac_i}
H_{vi} = && \eta_i \int d\omega_i (1-\eta_i |f_i(\omega_i)|^2)\nonumber\\
\ \ \ \ && \iint d\omega_{s1}' d\omega_{s2}' [h_{2i}^* (\omega_i ,\omega_{s1}' )h_{2i} (\omega_i ,\omega_{s2}' )s^* (\omega_{s2}' )s (\omega_{s1}' )].\end{eqnarray}
\end{subequations}
Accordingly, the normalized intensity noise of signal (idler) beam is
\begin{equation}
\label{RsAndRi}
R_j=\frac{ \Delta I_j^2 }{I_j}=\frac{ H_{j1}+H_{j2}+H_{vj} }{\eta_j \int d\omega_j |f_j(\omega_j )|^2 S_j(\omega_j)}~~(j=s,i).
\end{equation}
Moreover, we have $SNR_{out}= \frac{I_s^2}{R_s}=\frac{|\alpha|^2 \eta_s ^2 (\int d\omega_s |f_s(\omega_s )|^2 S_s(\omega_s))^2}{H_{s1}+H_{s2}+H_{vs}}$ by substituting Eqs. (\ref{eq:signal_idler_Average Photon Number}) and (\ref{RsAndRi}) into Eq. (\ref{SNR-out}). According to the definition in Eq. (\ref{eq:NF1}), we obtain the formula of noise figure
\begin{equation}
\label{NF2}
NF=\frac{H_{s1}+H_{s2}+H_{vs}}{\eta_s ^2 (\int d\omega_s |f_s(\omega_s )|^2 S_s(\omega_s))^2}.
\end{equation}

Eqs. (\ref{eq:Photon_number_fluctuation_FWM_OPA})-(\ref{RsAndRi}) indicate that the intensity noises of the individual beams $\Delta I_j^2 $ are not only related to detection efficiencies, $\eta_s$ and $\eta_i$, but also to the spectral functions of filters, $f_s(\omega_s)$ and $f_i(\omega_i)$. When $\eta_s$ and $\eta_i$ are very low, i.e., $\eta_j \ll 1$ ($j=s,i$), the intensity fluctuation $\Delta I_{s(i)}^2$ in Eq.
(\ref{eq:Photon_number_fluctuation_FWM_OPA}) is dominated by the term $H_{vj}$ ($j=s,i$), while the contribution of the term $H_{jk}$ ($k=1,2 \ j=s,i$) is negligible. On the other hand, when the bandwidth of the filters $F_s$ and $F_i$ are much narrower than that of input signal, $\Delta I_{s(i)}^2$ is also dominated by the term  $H_{vj}$ ($j=s,i$). Since the non-ideal detector induces the vacuum noise to the individual signal or idler field, the former is easy to understand. However, the understanding of the latter is not so straightforward. We think this is because, when the signal and idler beams are in multi-frequency mode, the filter $F_s$ ($F_i$) with bandwidth narrower than that of the amplified signal (generated idler) beam can also be viewed as a loss, which introduce vacuum noise to the signal (idler) beam as well. Hence, the term $H_{vj}$ ($j=s,i$) in Eqs. (\ref{eq:H_vac_s}) and (\ref{eq:H_vac_i}) is originated from the loss induced vacuum. If the detection and collection efficiencies of the signal and idler beam are ideal, we would have $H_{vj}=0$ ($j=s,i$).

Finally, we derive the expression of the intensity difference noise of the twin beams by using the definition
\begin{equation}
\label{difference-fluctuation_general}
\Delta I_t ^2 = \langle I_t ^2 \rangle-\langle \hat{I}_t \rangle^2
 =\iint \langle \Psi | \hat{I}_t(t_1)\hat{I}_t(t_2) | \Psi \rangle   dt_1 dt_2- (\int \langle \Psi | I_t(t) | \Psi \rangle dt) ^2,
\end{equation}
where the operator $\hat{I}_t= \hat{I}_{s}- r \hat{I}_{i} $ is the photon number difference with a weight factor of AC response ratio $r$. Using Eqs. (\ref{eq:multimode_trans}), (\ref{eq:vacuum_mixing}), (\ref{I_j})-(\ref{field-operator}) and (\ref{eq:fluctuation_general}), Eq. (\ref{difference-fluctuation_general}) is transformed into
 \begin{eqnarray}
 \label{eq:It_H}
 \Delta I_t^2=&&[\langle \hat{I}_s \hat{I}_s \rangle +r^2\langle \hat{I}_i \hat{I}_i \rangle - 2r\langle \hat{I}_s \hat{I}_i \rangle] - (I_s-rI_i)^2\nonumber\\
=&&\Delta I_s^2+r^2\Delta I_i^2-2r(H_{si1}+H_{si2})|\alpha|^2,
 \end{eqnarray}
where
 \begin{eqnarray}
\label{eq:H_si1}
H_{si1} =&&\eta_s \eta_i \iint\ldots\int d\omega_s d\omega_i d\omega_{s1}'d\omega_{s2}'d\omega_{s3}'\nonumber\\&&\{|f_s(\omega_s)|^2|f_i(\omega_i)|^2 h_{1s}(\omega_s,\omega_{s1}') h_{2i}(\omega_i,\omega_{s1}')\nonumber\\ &&h^*_{1s}(\omega_s,\omega_{s2}')s^*(\omega_{s2}')h^*_{2i}(\omega_i,\omega_{s3}')s(\omega_{s3}')\}
\end{eqnarray}
and
\begin{eqnarray}
\label{eq:H_si2}
H_{si2} =&&\eta_s \eta_i \iint\ldots\int d\omega_s d\omega_i d\omega_i'd\omega_{s1}'d\omega_{s2}'\nonumber\\
&&\{|f_s(\omega_s)|^2|f_i(\omega_i)|^2 h^*_{2s}(\omega_s,\omega_i') h^*_{1i}(\omega_i,\omega_i')\nonumber\\ &&h_{1s}(\omega_s,\omega_{s1}')f(\omega_{s1}')h_{2i}(\omega_i,\omega_{s2}')f^*(\omega_{s2}')\}
\end{eqnarray}
are positive terms originated from the quantum correlation between signal and idler twin beams. It is obvious that $\Delta I_t^2$ is less than the sum of intensity noise of individual beams $\langle \hat{I}_s \hat{I}_s \rangle +r^2\langle \hat{I}_i \hat{I}_i \rangle$ due to the correlation of twin beams.
Consequently, similar to Eq. (\ref{eq:electrical_gain}), the general expression of the normalized intensity difference noise for pulsed twin beams
\begin{equation}
\label{eq:R_t2}
R_t=\frac{ [H_{s1}+H_{s2}+H_{vs}+r^2(H_{i1}+H_{i2}+H_{vi})-2r(H_{si1}+H_{si2})] |\alpha|^2} { I_s + r^2 I_i }
\end{equation}
is obtained by substituting Eqs. (\ref{eq:Photon_number_fluctuation_FWM_OPA}) and (\ref{eq:H_si1})-(\ref{eq:H_si2}) into Eq. (\ref{eq:It_H}). In addition, $R_t$ can be further minimized if the AC response ratio $r$ takes the optimized value
\begin{eqnarray}
r=r_{opt}=&&\frac{ \sqrt{4I_sI_i(H_{si1}+H_{si2})^2+[I_s(H_{i1}+H_{i2}+H_{vi})-I_i(H_{s1}+H_{s2}+H_{vs})]^2} }{2I_i(H_{si1}+H_{si2})}\nonumber\\
&&+\frac{I_i(H_{s1}+H_{s2}+H_{vs})-I_s(H_{i1}+H_{i2}+H{vi})}{2I_i(H_{si1}+H_{si2})}.
\end{eqnarray}

From the formulas of the photon number gain, noise figure and intensity difference noise in Eqs. (\ref{gDef}), (\ref{NF2}) and (\ref{eq:R_t2}), one sees that the key parameters of the FOPA are determined by the transformation functions in Eq. (\ref{eq:Trans_Intergral}), which highly depend on the JSF in Eq. (\ref{JSF_def}). In the following sections, we will focus on studying the noise characteristics of the pulse pumped FOPAs with two typical kinds of JSF.

\section{\label{Fac_JSF} Pulse pumped FOPA with spectrally factorable JSF:\protect\\}

We first study the FOPA with a spectrally factorable JSF. Under this condition, both the signal and idler modes are  in single temporal mode, so we expect to recover the results for single mode FOPA in Sec. \ref{SINGLE_MODE}. For the sake of brevity, in this section, we assume the quantum efficiencies of D1 and D2 are perfect, and the collection efficiency of the twin beams is ideal.

A factorable JSF, obtained by properly regulating the pump pulses and tailoring the dispersion of optical fibers, is written as~\cite{Palmett07}
\begin{equation}
\label{eq:Factotized JSF}
\psi^{(f)}(\omega_s' ,\omega_i')=\phi_s(\omega_s') \ast \varphi_i(\omega_i'),
\end{equation}
where $\phi_s(\omega_s')$ and $\varphi_i(\omega_i')$ with the normalization condition of $\int|\phi_{s}(\omega_{s}')|^2 d\omega_{s}'=1$ and $\int|\varphi_{i}(\omega_{i}')|^2 d\omega_{i}'=1$ are the probability amplitude of finding a pairs of  signal and idler photons within the frequency range of $\omega_s'\rightarrow \omega_s'+d\omega_s'$ and $\omega_i'\rightarrow \omega_i'+d\omega_i'$, respectively. In fact, looking at Eq. (\ref{eq:Factotized JSF}) from another point of view, we find $\phi_s(\omega_s')$ and $\varphi_i(\omega_i')$ can also be viewed as the gain spectra in signal and idler fields, respectively.

With the JSF in Eq. (\ref{eq:Factotized JSF}), the transformation functions in Eq. (\ref{eq:Trans_Intergral}) are simplified to
\begin{subequations}
\label{eq:Fac_trans_func}
\begin{eqnarray}
&&h_{1s}^{(f)}(\omega_s,\omega_s')= \delta(\omega_s-\omega_s')+( \cosh G -1 )\cdot \phi_s(\omega_s)\phi_s^*(\omega_s')\\
&&h_{2s}^{(f)}(\omega_s,\omega_i')= \sinh G \cdot \phi_s(\omega_s)\varphi_i(\omega_i')\\
&&h_{1i}^{(f)}(\omega_i,\omega_i')= \delta(\omega_i-\omega_i')+( \cosh G -1 )\cdot \varphi_i(\omega_i)\varphi_i^*(\omega_i')\\
&&h_{2i}^{(f)}(\omega_i,\omega_s')= \sinh G \cdot \varphi_i(\omega_i)\phi_s(\omega_s').
\end{eqnarray}
\end{subequations}
Note that in this section, we will use the superscript ${(f)}$ to represent the result of an FOPA with the factorable JSF.

With the JSF in Eq. (\ref{eq:Factotized JSF}), the average photon numbers of the amplified signal and generated idler beams per pulse in Eq.(\ref{eq:signal_idler_Average Photon Number}) are simplified to
\begin{subequations}
\label{eq:factorized_signal_avg_ph_number}
\begin{eqnarray}
&&I_s^{(f)} = \big( 1+| \emph{F} \cdot \sinh G|^2 \big) |\alpha|^2 \\
&&I_i^{(f)} = \big( |\emph{F} \cdot \sinh G|^2 \big) |\alpha|^2 ,
\end{eqnarray}
\end{subequations}
where the coefficient $\emph{F}=\int s(\omega_s) \phi_s^*(\omega_s) d\omega_s$ is determined by the matching between the spectra of signal injection $s(\omega)$ (see Eq. (\ref{eq:Coh_State})) and that of the gain in signal band $\phi_s^*(\omega)$. Therefore, we obtain the photon number gain of the FOPA \begin{equation}
\label{gainJSF}
g^{(f)}=1+| \emph{F} \cdot \sinh G|^2
\end{equation}
by substituting Eqs. (\ref{eq:factorized_signal_avg_ph_number}) and (\ref{input-sig}) into Eq. (\ref{Def_g}).

To show the factors influencing the noise performance of the FOPA with factorable JSF, we first substitute the simplified transformation function in Eq. (\ref{eq:Fac_trans_func}) into Eqs.(\ref{eq45})-(\ref{eq46}) and (\ref{eq:H_si1})-(\ref{eq:H_si2}). For the case of $\eta_s=\eta_i=|f_s(\omega)|=|f_i(\omega)|=1$, the expressions of the terms in Eqs. (\ref{eq45})-(\ref{eq46}) and (\ref{eq:H_si1})-(\ref{eq:H_si2}), which determine $\Delta I_j ^2$ ($j=s,i$) in Eq. (\ref{eq:Photon_number_fluctuation_FWM_OPA}) and $\Delta I_t ^2$ in Eq. (\ref{eq:It_H}), are simplified to
\begin{eqnarray}
\label{eq:H_s1_fac}
&&H^{(f)}_{s1} =1+|F|^2\cdot(|\cosh{G}|^4-1),
\end{eqnarray}
\begin{eqnarray}
\label{eq:H_i1_fac}
&&H^{(f)}_{i1} = |F|^2 \cdot |\sinh G|^4,
\end{eqnarray}
\begin{eqnarray}
&&H^{(f)}_{s2}=H^{(f)}_{i2}= H^{(f)}_{si1}=H^{(f)}_{si2}= |F \cdot \sinh G  \cosh G|^2,
\end{eqnarray}
\begin{eqnarray}
H^{(f)}_{vj}=0.
\end{eqnarray}
Consequently, the normalized intensity noise of individual signal and idler beams are
\begin{equation}
\label{Rs_Fac}
R^{(f)}_s=\frac{1+|F|^2(|\sinh G|^4+|\sinh G \cosh G|^2-1)}{1+|F\cdot\sinh G|^2}
\end{equation}
and
\begin{equation}
\label{Ri_Fac}
R^{(f)}_i=|\sinh G|^2+|\cosh G|^2,
\end{equation}
respectively. Moreover, according to Eq. (\ref{eq:factorized_signal_avg_ph_number}), the noise figure in Eq. (\ref{NF2}) and the normalized intensity difference noise of twin beams in Eq. (\ref{eq:R_t2}) can be respectively simplified to
\begin{equation}
\label{eq:NF_Fac}
NF^{(f)}=\frac{2}{|F|^2}
\end{equation}
and
\begin{equation}
 \label{eq:Rt_Fac}
 R^{(f)}_t=\frac{1}{2 |F \cdot \sinh{G}|^2 + 1}\ \ (r=1).
 \end{equation}

Equations (\ref{Rs_Fac})-(\ref{eq:Rt_Fac}) show that except for the normalized intensity noise of idler beam $R^{(f)}_i$, the remaining three parameters, $R^{(f)}_s$, $NF^{(f)}$ and $R^{(f)}_t$, are associated with the matching coefficient $F$. For the case of
 $F=1$, if we respectively replace the terms $1+|\sinh G|^2$ and $|\sinh G|^2$ with $|\mu|^2$ and $|\nu|^2$, the analytical expressions of the key parameters (Eqs. (\ref{Rs_Fac})-(\ref{eq:Rt_Fac})) will be the same as those of the single frequency mode FOPA in Sec. \ref{SINGLE_MODE}. The results indicate that our multi-mode theory of the pulse pumped FOPA are valid.

An marked difference between the FOPAs described by single temporal mode and single frequency mode is the spectra of signal and idler beams. The spectra of the amplified signal and generated idler beams obtained by substituting Eq. (\ref{eq:Fac_trans_func}) into Eq.(\ref{spectra}) are written as
\begin{subequations}
\label{spectrum_fa}
\begin{eqnarray}
S_{s}^{(f)}(\omega_s)=&&|s(\omega_s)|^2+|F\cdot \phi_s(\omega_s)|^2|\cosh G-1|^2
\nonumber\\&&+F^*(\cosh G-1)\phi_s(\omega_s)\phi_s^*(\omega_s)+F(\cosh G-1)^*\phi_s^*(\omega_s)\phi_s(\omega_s)
\end{eqnarray}
and
\begin{eqnarray}
S_{i}^{(f)}(\omega_i)=&&|\sinh G|^2|\varphi_i(\omega_i)|^2,
\end{eqnarray}
\end{subequations}
respectively. From Eq. (\ref{spectrum_fa}), one sees that the spectrum of generated idler is always the same as the spectrum of the gain in idler field, however, different from the FOPA pumped with a single frequency laser, the spectra of the injected signal and amplified signal beams are different unless the condition $F=1$ is fulfilled.

To further understand the difference between the FOPAs described by single temporal mode and single frequency mode, we then compare the noise performance of the two cases. We find that for a given value of $G$ (or $g$), the noise figure and the intensity difference noise of the twin beams in Eqs. (\ref{eq:NF_Fac}) and (\ref{eq:Rt_Fac}) are worse than that predicted by Eqs (\ref{NF_s_sm}) and (\ref{R_t_sm}) unless $F=1$. In practice, it is very difficult to maintain the spectral matching condition $\phi_s(\omega_s) = s(\omega_s)$ in the high gain regime due to the self-phase modulation and cross-phase modulation induced spectral broadening~\cite{Cui2012}. Hence, it is very challenging to experimentally realize a pulse pumped FOPA capable of being described by a single mode theory.

\section{\label{perfect_phase_matched_JSF}Pulse pumped FOPA with spectrally non-factorable JSF:\protect\\}

For the conventional optical communication systems, an FOPA with an extremely broad gain bandwidth is desirable.
When the gain bandwidth of the FOPA is much broader than the bandwidths of the pump and input signal, the JSF can be simplified by assuming the perfect phase matching condition $\Delta k=0$. In this case, the JSF in Eq. (\ref{JSF_def}) is rewritten as~\cite{Yang2011}
\begin{equation}
\label{eq:JSF_PM}
\psi^{(b)}(\omega_s',\omega_i')=\frac{C}{2\sqrt{\pi}\sigma_p}\exp{\{\frac{-(\omega_s'+\omega_i'-2\omega_{po})^2}{4\sigma_p^2}\}},
\end{equation}
which is obviously non-factorable. In this section, we focus on numerical investigation of the quantum noise performance of this kind of FOPA.

With the JSF in Eq. (\ref{eq:JSF_PM}), the transformation functions in Eq.(\ref{eq:Trans_Intergral}) are reformulated as
\begin{subequations}
\label{Trans_unFac}
\begin{eqnarray}
&&h_{1s}^{(b)}(\omega,\omega')=h_{1i}^{(b)}(\omega,\omega')\nonumber\\&&= \delta(\omega-\omega')+ \sum_{n=1}^{\infty}
\frac{1}{\sqrt{2n}(2n)!} \frac{G'^{2n}}{2\sqrt{\pi}\sigma_p}  \textit{e}^{  \frac{-(\omega-\omega')^2}{4\sigma_p^2 \cdot 2n}  } ~~\label{eq:PM_JSF_h1} \\
&&h_{2s}^{(b)}(\omega,\omega')=h_{2i}^{(b)}(\omega,\omega')\nonumber\\&&
=\sum_{n=0}^{\infty}
\frac{1}{\sqrt{2n+1}(2n+1)!} \frac{G'^{2n+1}}{2\sqrt{\pi}\sigma_p} \textit{e}^{  \frac{-(\omega+\omega'-2\omega_{po})^2}{4\sigma_p^2 \cdot (2n+1)}  },
~~\label{eq:PM_JSF_h2}
\end{eqnarray}
\end{subequations}
where $G'=C G $ , and the superscript ${(b)}$ represents the result for the FOPA with broad gain bandwidth. For the sake of convenience, the frequency variables $\omega_{s(i)}$ and $\omega_{s(i)}'$ used in previous sections will be replaced with $\omega$ and $\omega'$ hereinafter.

\subsection{\label{pmf_Filtered} Key parameters of the FOPA in the ideal conditions}

We first study the factors influencing the key parameters of the FOPA by assuming the detectors are perfect, and collection efficiency of the twin beams is ideal, i.e., $\eta_s=\eta_i=1$ and $|f_s(\omega)|=|f_i(\omega)|=1$.
Without loss of the generality, we assume the spectral function of the pulsed input signal can be described by the Gaussian function
\begin{equation}
\label{eq:guassian_input_signal_spectrum}
s(\omega)=\frac{1}{\sqrt{\pi^{1/2} \sigma} }\exp{\{\frac{-(\omega-\omega_{s0})^2}{2\sigma^2}\}},
\end{equation}
where $\sigma$ denotes the bandwidth. From Eqs.  (\ref{Trans_unFac}) and (\ref{eq:guassian_input_signal_spectrum}), we can rewrite the power spectra of the amplified signal and generated idler beams in Eq. (\ref{spectra}) as
\begin{subequations}
\label{spectra_unFac}
\begin{eqnarray}
&&S^{(b)}_s(\omega)= \frac{1}{\sqrt{\pi}} \sum_{n_1=0}^{\infty} \sum_{n_2=0}^{\infty} \frac{|G'|^{2n_1+2n_2}}{(2n_1)!(2n_2)!} \exp{\{\frac{-(\omega-\omega_{s0})^2}{2(\sigma^2+4n_1\sigma_p^2)}\}}
\nonumber \\&&\ \ \ \ \ \times \exp{\{\frac{-(\omega-\omega_{s0})^2}{2(\sigma^2+4n_2\sigma_p^2)}\}}\sqrt{\frac{\sigma^2}{(\sigma^2+4n_1\sigma_p^2) (\sigma^2+4n_2\sigma_p^2)}},
\\
&&S^{(b)}_i(\omega)=  \frac{1}{\sqrt{\pi}}  \sum_{n_1=0}^{\infty} \sum_{n_2=0}^{\infty} \frac{|G'|^{2n_1+2n_2+2}}{(2n_1+1)!(2n_2+1)!} \exp{\{\frac{-(\omega-\omega_{i0})^2}{2(\sigma^2+2(2n_1+1)\sigma_p^2)}\}}
\nonumber \\&&\ \ \ \ \ \ \times \exp{\{\frac{-(\omega-\omega_{i0})^2}{2(\sigma^2+2(2n_2+1)\sigma_p^2)}\}}\sqrt{\frac{\sigma^2}{(\sigma^2+2(2n_1+1)\sigma_p^2) (\sigma^2+2(2n_2+1)\sigma_p^2)}}.
\end{eqnarray}
\end{subequations}
For all the terms of the infinite series in Eqs. (\ref{spectra_unFac}a) and (\ref{spectra_unFac}b), whose order corresponds to $n_1 \not= 0$ or $n_2 \not= 0$, their bandwidths are larger than that of the input power spectrum $|s(\omega_s)|^2$. Moreover, since the bandwidth of the term in infinite series increases with the order determined by the integers $n_1$ and $n_2$, the bandwidths of both signal and idler twin beams increase with $G'$.

Using Eq. (\ref{spectra_unFac}),  the average photon numbers of amplified signal field and generated idler beams in Eq. (\ref{eq:signal_idler_Average Photon Number}) can be rewritten as
\begin{subequations}
\label{I_unFac}
\begin{eqnarray}
&&I^{(b)}_s=|\alpha|^2\sum_{n_1=0}^{\infty} \sum_{n_2=0}^{\infty} \frac{|G'|^{2n_1+2n_2}}{(2n_1)!(2n_2)!}\sqrt{\frac{1}{1+2(n_1+n_2)p^2}}
\\
&&I^{(b)}_i=|\alpha|^2\sum_{n_1=0}^{\infty} \sum_{n_2=0}^{\infty} \frac{|G'|^{2n_1+2n_2+2}}{(2n_1+1)!(2n_2+1)!}\sqrt{\frac{1}{1+2(n_1+n_2+1)p^2}},
\end{eqnarray}
\end{subequations}
where $p=\sigma_p/\sigma$ describes the ratio of the pump bandwidth to the input signal bandwidth.

For all the numerically simulated results presented hereinafter, the coefficient $G'$ is within the range of $0<|G'|<4$, and the series in Eqs. (\ref{Trans_unFac}), (\ref{spectra_unFac}) and (\ref{I_unFac}) are calculated up to the order of $n_1=10$ and $n_2=10$. To ensure the correctness of this truncation approximation,
we numerically verify the general relation $\frac{I_s^{(b)}-I_i^{(b)}}{I_{in}}=1$ ($I_{in}=|\alpha|^2$), which is the inherent nature of signal and idler twin beams originated from the energy conservation of FWM. After calculating the series in Eq. (\ref{I_unFac}) to the order of $n_1=10$ and $n_2=10$, we find that the difference between the calculated result of $\frac{I_s^{(b)}-I_i^{(b)}}{|\alpha|^2}$ and ``1" is less than $10^{-11}$, showing the validity of the approximation.

\begin{figure}[h]
 \includegraphics[width=0.48\textwidth]{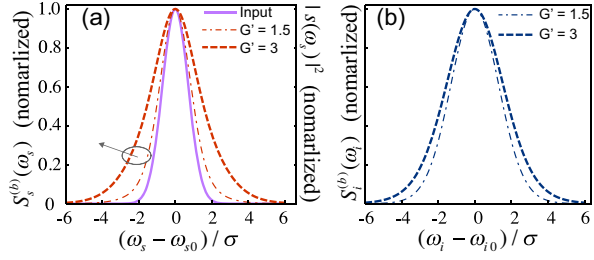}               %Here is how to import EPS art
\caption{\label{spectrum} The normalized power spectra of  (a) amplifed signal and (b) generated idler beams  at different $G'$ under the condition $\sigma=\sigma_p$ . As a comparison, the power spetrum of input signal  $|s(\omega_s)|^2$ is plotted in (a) as well.}
\end{figure}

Figures \ref{spectrum}(a) and \ref{spectrum}(b) show the normalized power spectra of the amplified signal and generated idler beams. In the plots,   Eqs. (\ref{spectra_unFac}a) and (\ref{spectra_unFac}b) are calculated by varying the value of $G'$ under the condition $\sigma=\sigma_p$. As a comparison, we also show the power spectrum of input signal $|s(\omega)|^2$ in Fig. \ref{spectrum}(a). It is obvious that the bandwidths of both signal and idler twin beams are broader than  that of the input signal  and they increase with $G'$, showing the spectrum broadening effect.  Moreover, in the low gain regime, the spectrum of amplified signal is mainly determined by the input signal, so  the bandwidth of generated idler beam, which is the convolution of the spectra of the input signal and pump, is greater than that of the signal beam. However, in the high gain regime, the bandwidths of signal and idler twin beams are about equal because both the spectra of signal and idler beams are the convolution of the spectra of pump and its counterparts.

\begin{figure}[h]
 \includegraphics[width=0.33\textwidth]{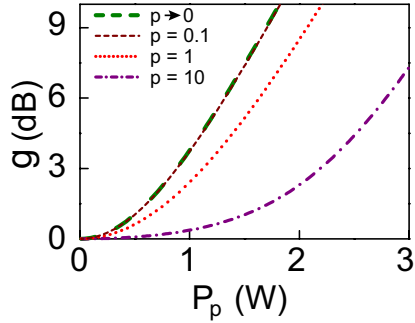}               %Here is how to import EPS art
\caption{\label{fig:Photon_Number_gain_no_filtered} Photon number gain $g$ versus the peak pump power for the different ratio $p=\sigma_p/\sigma$. In the calculation, $\Delta k=0$, $\eta_s=\eta_i=|f_s(\omega_s)|=|f_i(\omega_i)|=1$, the parameter $\beta=\frac{-8 \emph{i} C_1 \gamma L c A_{eff}^2}{3 \hbar \omega_{p0}} =1$, and $G' = \beta P_p$.}
\end{figure}

Figure \ref{fig:Photon_Number_gain_no_filtered} plots the dependence of the photon number gain $g=I_s^{(b)}/|\alpha|^2$ upon the pump power for the different ratio $p=\sigma_p/\sigma$. In the calculation, $G' \propto P_p$ is changed by varying the peak pump power $P_p$. One sees that at a fixed peak power $P_p$, $g$ increases with the decrease of $p$ because the overlap of temporal mode, which is required for maximizing the gain of FWM, is improved by decreasing the bandwidth of pump. When $p$ is less than $0.1$, further decreasing the ratio to $p\rightarrow 0$ does not result in an obvious increase in $g$, because no more space is left for improving the temporal mode overlapping. The result in Fig. \ref{fig:Photon_Number_gain_no_filtered} implies the pump with a longer pulse duration (narrower bandwidth) gives a better mode overlapping. However, it is worth noting that the pump with a short pulse duration helps to obtain a high gain under the condition of low average pump power, which is of practical importance.

%To show the factors influencing the noise performance of the FOPA with factorable JSF, we first substitute the simplified transformation function in Eq. (\ref{eq:Fac_trans_func}) into Eqs.(\ref{eq:H_s1})-(\ref{NF2}). For the case of $\eta_s=\eta_i=|f_s(\omega)|=|f_i(\omega)|=1$, the expressions of terms in Eqs. (\ref{eq:H_s1})-(\ref{eq:H_vac_i}) and (\ref{eq:H_si1})-(\ref{eq:H_si2}), which determine $\Delta I_j ^2$ ($j=s,i$) in Eq. (\ref{eq:Photon_number_fluctuation_FWM_OPA}) and $\Delta I_t ^2$ in Eq. (\ref{eq:It_H}), are simplified to

To investigate the quantum noise performance of the FOPA with broad gain bandwidth, we first deduce the formulas of terms $H_{jk}$ ($j=s,i$, $k=1,2$) and $H_{vj}$ ($j=s,i$), which determine $\Delta I_j ^2$ ($j=s,i$) in Eq. (\ref{eq:Photon_number_fluctuation_FWM_OPA}). By substituting Eqs. (\ref{Trans_unFac}) and (\ref{eq:guassian_input_signal_spectrum}) into Eqs. (\ref{eq45})-(\ref{eq46}), we obtain
 \begin{eqnarray}
 \label{eq:Hs1_pm}
  H_{s1}^{(b)}=&& \sum_{n_1 \rightarrow n_4}^{\infty} \frac{|G'|^{2n_1+2n_2+2n_3+2n_4}}{(2n_1)!(2n_2)!(2n_3)!(2n_4)!}
 \nonumber\\&& \ \ \ \  \sqrt{\frac{1}{1+2(n_1+n_2+n_3+n_4)p^2}}
 \end{eqnarray}
 \begin{eqnarray}
 \label{eq:Hs2_pm}
 H^{(b)}_{s2}=&&H^{(b)}_{i2}
 \nonumber\\
 =&&\sum_{n_1 \rightarrow n_4}^{\infty} \frac{|G'|^{2n_1+2n_2+2n_3+2n_4+2}}{(2n_1)!(2n_2)!(2n_3+1)!(2n_4+1)!}
 \nonumber\\&& \ \ \ \  \sqrt{\frac{1}{1+2(n_1+n_2+n_3+n_4+1)p^2}}
 \end{eqnarray}
  \begin{eqnarray}
   \label{eq:Hi1_pm}
H^{(b)}_{i1}= &&\sum_{n_1 \rightarrow n_4}^{\infty} \frac{|G'|^{2n_1+2n_2+2n_3+2n_4+4}}{(2n_1+1)!(2n_2+1)!(2n_3+1)!(2n_4+1)!}
 \nonumber\\&& \ \ \ \  \sqrt{\frac{1}{1+2(n_1+n_2+n_3+n_4+2)p^2}}
 \end{eqnarray}
 \begin{eqnarray}
 \label{eq:Hsivac_pm}
 &&H_{vj}=0~~(j=s,i).
 \end{eqnarray}

\begin{figure}[h]
\includegraphics[width=0.33\textwidth]{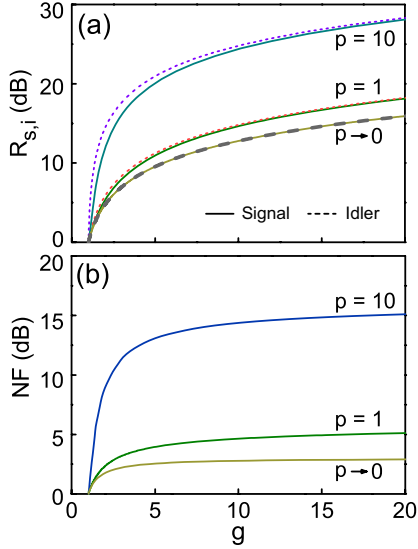}               %Here is how to import EPS art
\caption{\label{fig:relative_intensity_noise_no_filtered}(a) The normalized intensity noise of the individual signal/ idler beams $R_{j}$ ($j=s,i$) and (b) noise figure NF versus the photon number gain $g$ for the different ratio $p=\sigma_p/\sigma$. In this calculation, $\Delta k=0$, and $\eta_s=\eta_i=|f_s(\omega)|=|f_i(\omega)|=1$.}
\end{figure}

With the above, we then numerically study the dependence of the normalized intensity noise of individual beams $R_j=\Delta I_j^2/I_j$ ($j=s,i$) upon the ratio $p$.
Figure \ref{fig:relative_intensity_noise_no_filtered}(a) plots $R_s$ and $R_i$ versus $g$ for the cases of $p = 10$,
$p = 1$, and $p \rightarrow 0$, respectively. From Fig. \ref{fig:relative_intensity_noise_no_filtered}(a), one sees that for a fixed value of $g$ ($p$), both $R_s$ and $R_i$ increase with $p$ ($g$). Moreover, for the given values of $p$ and $g$, $R_i$ is greater than $R_s$. This is different from the results of $R_j^{(s)}$  ($j=s,i$) for a single mode FOPA [see Eq. (\ref{R_CW})], but has been experimentally confirmed in references \cite{Sharping01} and \cite{Guo12}.

We also compute the noise figure $NF$ in Eq. (\ref{NF2}).
Figure \ref{fig:relative_intensity_noise_no_filtered}(b) plots $NF$ versus $g$ for $p = 10$,
$p = 1$, and $p \rightarrow 0$, respectively. One sees that for a fixed ratio $p$, $NF$ increases with $g$ in the low gain regime, and approaches a constant in the high gain regime. While for a fixed $g$, the value of NF increases with the ratio $p$. It is obvious that for the pulsed pump case with $p \neq 0$, $NF$ in the high gain regime is always greater than the well known 3dB-limit of a single mode FOPA.

 \begin{figure}[]
\includegraphics[width=0.55\textwidth]{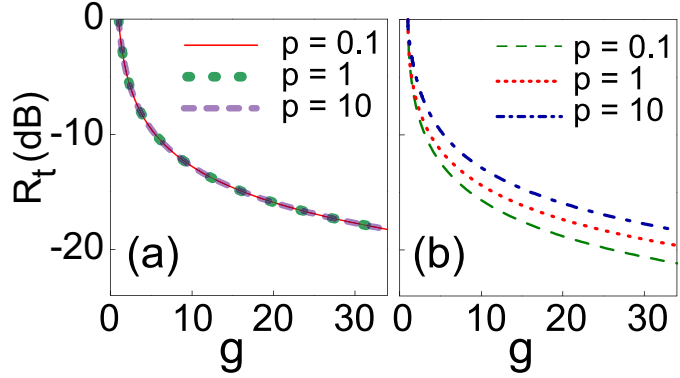}
\caption{\label{fig:twin_beams_no_filtered} The normalized intensity difference noise of the twin beams $R_t$ versus photon number gain $g$ for $p=0.1$, $p=1$ and $p=10$ when $r$ is set to (a) $r=1$ and (b) $r=r_{opt}$, respectively.}
\end{figure}

 \begin{figure}[]
\includegraphics[width=0.4\textwidth]{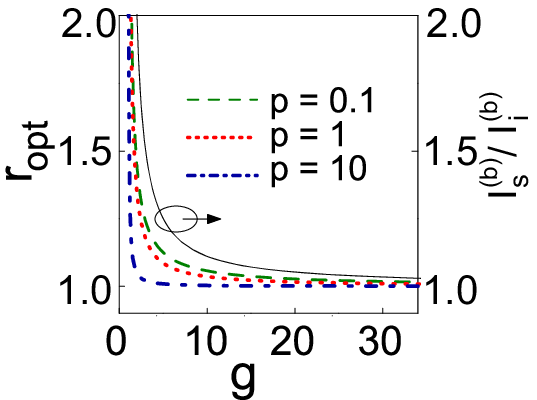}
\caption{\label{fig6} The optimized value $r_{opt}$ and the photon number ratio $I_s^{(b)}/I_i^{(b)}$ as a function of $g$. In this calculation, the phase mismatching term of the FOPA is $\Delta k=0$, and the detection efficiencies and the filters satisfy $\eta_s=\eta_i=|f_s(\omega)|=|f_i(\omega)|=1$.}
\end{figure}
To study the factors influencing the normalized intensity difference noise of the twin beams $R_t$ in Eq. (\ref{eq:R_t2}), we need to calculate $H^{(b)}_{si1}$ and $H^{(b)}_{si2}$ as well. By substituting Eqs. (\ref{Trans_unFac}) and (\ref{eq:guassian_input_signal_spectrum}) into Eqs. (\ref{eq:H_si1}) and (\ref{eq:H_si2}), we find the following relation
 \begin{equation}
 H^{(b)}_{si1}=H^{(b)}_{si2} =H^{(b)}_{s2}=H^{(b)}_{i2}.
 \end{equation}
In Figs. \ref{fig:twin_beams_no_filtered}(a) and \ref{fig:twin_beams_no_filtered}(b), we plot $R_t$ as a function of $g$ for different values of $p$ when the AC response ratio $r$ takes the values of $r=1$ and $r=r_{opt}$, respectively. Fig. \ref{fig:twin_beams_no_filtered}(a) shows $R_t$ is independent on $p$ and always decreases with the increase of $g$; while Fig. \ref{fig:twin_beams_no_filtered}(b) demonstrates that $R_t$ still decreases with the increase of $g$, but $R_t$ can be further decreased and depends upon the ratio $p$. For a fixed gain $g$, $R_t$ decreases with $p$, which is different from the case of $r=1$. Moreover, when $p$ is very small, for example, $p=0.1$, the difference between the $R_t$-values with $r=1$ and $r=r_{opt}$ is about 3 dB in the high gain limit; when $p$ is very big, for example, $p=10$, $R_t$-values with $r=1$ and $r=r_{opt}$ are almost the same.

The results in Figs. \ref{fig:twin_beams_no_filtered}(a) and \ref{fig:twin_beams_no_filtered}(b) not only indicate the noise reduction of $R_t$ can be improved by properly reducing the ratio $p$, but also
imply that $r_{opt}$ might depend upon $p$ and $r_{opt}$ might be very close to 1 when the value of $p$ is greater than 10. This is shown in Fig. \ref{fig6}, in which we plot $r_{opt}$ as a function of $g$ for $p=0.1$, $p=1$ and $p=10$. In addition, we also plot the photon number ratio $I_s^{(b)}/I_i^{(b)}$ versus $g$ in Fig. \ref{fig6}. One sees that $r_{opt}$ is always within the range of $I_s^{(b)}/I_i^{(b)}$ and $1$. For a fixed value of $g$, $r_{opt}$ decreases with the increase of $p$. As we expected, $r_{opt}$ obtained for the case of $p=10$ is about 1 if $g$ is greater than $3$. Moreover, in the high gain limit, both $r_{opt}$ and $I_s^{(b)}/I_i^{(b)}$ approach $1$, which is irrelevant to $p$.

\subsection{\label{pmf_Filtered} Influence of the detection loss and collection loss of twin beams on $R_t$}

In practice, the ideal condition $\eta_s=\eta_i=|f_s(\omega)|=|f_i(\omega)|=1$ can never be fulfilled,
the measured intensity difference noise $R_t$ is closely related to the loss of the FOPA system. So we need to study the influence of the quantum efficiencies ($\eta_s$ and $\eta_i$) and collection efficiency of twin beams. In this subsection, we assume the filters placed at the output port of the FOPA (see Fig. 1) have a Gaussian shaped spectrum
\begin{equation}
\label{eq:filter_shape_Guassian}
f_j(\omega)=\exp{\{\frac{-(\omega-\omega_{jo})^2}{2\sigma_f^2}\}}\ \ \ (j=s,i),
\end{equation}
where $\omega_{jo}$ and $\sigma_f$ denote the central frequency and bandwidth of the filter $F_{j}$  $(j=s,i)$, respectively.

Using the transformation functions, input signal spectrum and the spectrum of $F_{j}$ $(j=s,i)$ in Eqs.  (\ref{Trans_unFac}), (\ref{eq:guassian_input_signal_spectrum}) and (\ref{eq:filter_shape_Guassian}), respectively, the measured photon numbers of the amplified signal and generated idler beams in Eq. (\ref{eq:signal_idler_Average Photon Number}) are rewritten as
\begin{subequations}
\label{eq:perfect_phase_matching_filtered_signal_ave_pho_num}
\begin{eqnarray}
&&I'^{(b)}_s=\eta_s \sum_{n_1=0}^{\infty} \sum_{n_2=0}^{\infty} \frac{|G'|^{2n_1+2n_2}}{(2n_1)!(2n_2)!}
\nonumber\\&&\ \ \ \ \  \{ \ s^2(4n_1 p^2+1)(4n_2 p^2+1)
\nonumber\\&&\ \ \ \ \ \ \  +[2(n_1+n_2)p^2+1]\  \}^{-1/2}
\\
&&I'^{(b)}_i=\eta_i \sum_{n_1=0}^{\infty} \sum_{n_2=0}^{\infty} \frac{|G'|^{2n_1+2n_2+2}}{(2n_1+1)!(2n_2+1)!}
\nonumber\\&&\ \ \ \ \ \{\ s^2[(4n_1+2)p^2+1][(4n_2+2)p^2+1]
\nonumber\\&&\ \ \ \ \ \ \ +[2(n_1+n_2+1)p^2+1]\  \}^{-1/2},
\end{eqnarray}
\end{subequations}
where $s=\sigma/\sigma_f$ is the bandwidth ratio of the input signal to filter $F_{s(i)}$. For clarity, we add an apostrophe to label the result of a broad-band FOPA with non-ideal collection and detection efficiencies. Moreover, by substituting Eqs.  (\ref{Trans_unFac}), (\ref{eq:guassian_input_signal_spectrum}) and (\ref{eq:filter_shape_Guassian}) into Eqs.(\ref{eq45})-(\ref{eq46}), the expressions of the terms of $\Delta I_j ^2$ ($j=s,i$) in Eq. (\ref{eq:Photon_number_fluctuation_FWM_OPA}) are reformulated as
\begin{eqnarray}
 \label{Hs1_f}
 &&H'^{(b)}_{s1} =\eta^2_s\sum_{n_1 \rightarrow n_4}^{\infty} \frac{|G'|^{2n_1+2n_2+2n_3+2n_4}}{(2n_1)!(2n_2)!(2n_3)!(2n_4)!}
 \nonumber\\&&\ \ \ \ \ \ \ \ \sqrt{\frac{1}{1+\xi^{f}_{2n_1,2n_2,2n_3,2n_4}}},
 \end{eqnarray}
 \begin{eqnarray}
 &&H'^{(b)}_{s2} =\eta^2_s\sum_{n_1 \rightarrow n_4}^{\infty} \frac{|G'|^{2n_1+2n_2+2n_3+2n_4+2}}{(2n_1)!(2n_2)!(2n_3+1)!(2n_4+1)!}
 \nonumber\\&&\ \ \ \ \ \ \ \ \sqrt{\frac{1}{1+\xi^{f}_{2n_1,2n_2,2n_3+1,2n_4+1}}},
 \end{eqnarray}
 \begin{eqnarray}
 &&H'^{(b)}_{i1} =\eta^2_i\sum_{n_1 \rightarrow n_4}^{\infty} \frac{|G'|^{2n_1+2n_2+2n_3+2n_4+4}}{(2n_1+1)!(2n_2+1)!(2n_3+1)!(2n_4+1)!}
 \nonumber\\&&\ \ \ \ \ \ \ \sqrt{\frac{1}{1+\xi^{f}_{2n_1+1,2n_2+1,2n_3+1,2n_4+1}}},
 \end{eqnarray}
 \begin{eqnarray}
  \label{Hi2_f}
 &&H'^{(b)}_{i2} =\eta^2_i\sum_{n_1 \rightarrow n_4}^{\infty} \frac{|G'|^{2n_1+2n_2+2n_3+2n_4+1}}{(2n_1+1)!(2n_2+1)!(2n_3)!(2n_4)!}
 \nonumber\\&&\ \ \ \ \ \ \ \ \sqrt{\frac{1}{1+\xi^{f}_{2n_1+1,2n_2+1,2n_3,2n_4}}},
 \end{eqnarray}
  \begin{eqnarray}
  \label{Hvacs_f}
 &&H'^{(b)}_{vs} = \eta_s \sum_{n_1=0}^{\infty} \sum_{n_2=0}^{\infty} \frac{|G'|^{2n_1+2n_2}}{(2n_1)!(2n_2)!}
 \nonumber\\&&\ \ \ \ \{\sqrt{\frac{1}{1+\xi^{vac}_{2n_1,2n_2,1}}}-\eta_s\sqrt{\frac{1}{1+\xi^{vac}_{2n_1,2n_2,2}}}\},
 \end{eqnarray}
  \begin{eqnarray}
 \label{Hvaci_f}
 &&H'^{(b)}_{vi} = \eta_i \sum_{n_1=0}^{\infty} \sum_{n_2=0}^{\infty} \frac{|G'|^{2n_1+2n_2+2}}{(2n_1+1)!(2n_2+1)!}
 \nonumber\\&&\ \ \{\sqrt{\frac{1}{1+\xi^{vac}_{2n_1+1,2n_2+1,1}}}-\eta_i\sqrt{\frac{1}{1+\xi^{vac}_{2n_1+1,2n_2+1,2}}}\}.
 \end{eqnarray}
 By substituting Eqs.  (\ref{Trans_unFac}), (\ref{eq:guassian_input_signal_spectrum}) and (\ref{eq:filter_shape_Guassian}) into Eqs. (\ref{eq:H_si1}) and (\ref{eq:H_si2}), the other two terms of $R_t$ in Eq. (\ref{eq:R_t2}) are reformulated as
  \begin{eqnarray}
  \label{Hsi1_f}
 &&H'^{(b)}_{si1} = \eta_s \eta_i \sum_{n_1 \rightarrow n_4}^{\infty} \frac{|G'|^{2n_1+2n_2+2n_3+2n_4+2}}{(2n_1)!(2n_2+1)!(2n_3)!(2n_4+1)!}
 \nonumber\\&&\ \ \ \ \ \ \ \ \sqrt{\frac{1}{1+\xi^{f}_{2n_1,2n_2+1,2n_3,2n_4+1}}}
 \end{eqnarray}
 \begin{eqnarray}
  \label{Hsi2_f}
 &&H'^{(b)}_{si2} = \eta_s \eta_i \sum_{n_1 \rightarrow n_4}^{\infty} \frac{|G'|^{2n_1+2n_2+2n_3+2n_4+2}}{(2n_1+1)!(2n_2)!(2n_3+1)!(2n_4)!}
 \nonumber\\&&\ \ \ \ \ \ \ \ \sqrt{\frac{1}{1+\xi^{f}_{2n_1+1,2n_2,2n_3+1,2n_4}}}.
 \end{eqnarray}
Here the coefficients
\begin{eqnarray}
&&\xi^{f}_{k_1,k_2,k_3,k_4}=(k_1+k_2+k_3+k_4)p^2
\nonumber\\&&\ \ \ \ \ \ \ \ \ \ +s^2 \{ 2+4(k_1+k_2+k_3+k_4)p^2+4 [k_2 (k_3+k_4)+k_1 (2 k_2+k_3+k_4)] p^4 \}
\nonumber\\&&\ \ \ \ \ \ \ \ \ \ +4 s^4  p^2 (1+2 k_1 p^2) (1+2 k_2 p^2) (k_3+k_4)
\\
&&\xi^{vac}_{k_1,k_2,x}=x\cdot(k_1+k_2)p^2
\nonumber\\&&\ \ \ \ \ \ \ \ \ \ +s^2[1+2k_1p^2][1+2k_2p^2]~~~(x=1,2)
\end{eqnarray}
are introduced to simplify the Eqs. (\ref{Hs1_f})-(\ref{Hsi2_f}).

For a fixed value of $p=\sigma_p /\sigma$, if the bandwidth of filters $F_{s(i)}$ is much narrower than that of the input signal, namely, $s=\sigma/\sigma_f\rightarrow \infty$, we have $\xi^{f}_{k_1,k_2,k_3,k_4} \gg \xi^{vac}_{k_1,k_2,x}$. In this case, the collection efficiency of the twin beams is very low due to the multi-mode nature of the signal and idler fields~\cite{Li2010}. Since the narrow band filters $F_{s(i)}$ introduce the vacuum noise into the individual signal and idler beams, for a given $G'$, the value of $H'_{vj}$ $(j=s,i)$ in Eqs (\ref{Hvacs_f})-(\ref{Hvaci_f}) approaches to $I'^{(b)}_j\ (j=s,i)$ and is dominant over the terms $H'_{jl}$ ($l=1,2$) and $H'_{sil}$ ($l=1,2$) (Eqs. (\ref{Hs1_f})-(\ref{Hi2_f}) and (\ref{Hsi1_f})-(\ref{Hsi2_f})). Hence, the normalized intensity noise and normalized intensity difference noise approach to the SNL, i.e. $R_j \rightarrow 1 \ (j=s,i)$ and $R_t \rightarrow 1$.

\begin{figure}[]
\includegraphics[width=0.85\textwidth]{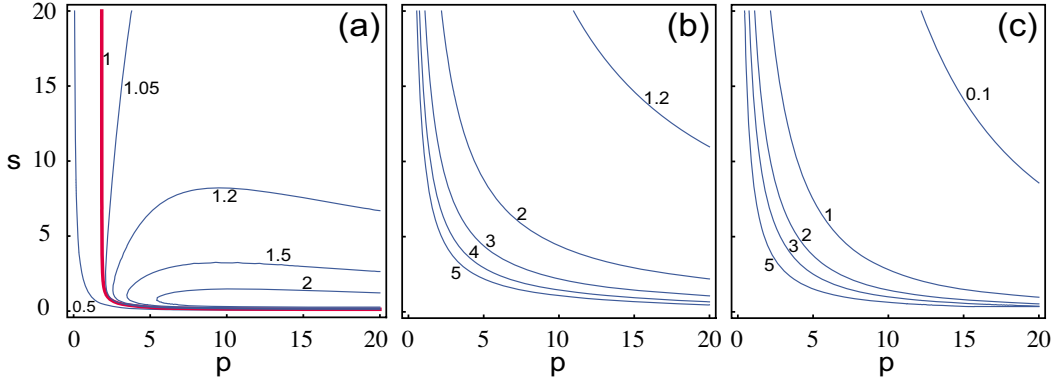}
 \caption{\label{Contour}  Contour plot of (a) the normalized intensity difference noise $R_t$, (b) $(\Delta I_s ^2+\Delta I_i ^2)/(I_s+I_i)$, and (c) $2(H_{si1}+H_{si2}) |\alpha|^2/(I_s+I_i)$ as a function of the ratios $p$ and $s$. In this calculation, $\eta_s = \eta_i = 1$, $G'=3$, and $r=1$, and $R_t=\frac {\Delta I_s ^2+\Delta I_i ^2-2(H_{si1}+H_{si2})|\alpha|^2}{I_s+I_i}$ }
 \end{figure}

In general, the collection efficiency of the twin beams is not only associated with the ratio $s$, but also depends on the ratio $p$. To reduce the detrimental effect of the collection efficiency on $R_t$, both $s$ and $p$ should be small enough unless the condition $s\rightarrow 0$ or $p\rightarrow 0$ is satisfied~\cite{Li2010}. To illustrate this point, we plot the contour of the normalized intensity difference noise $R_t$ as a function of $p$ and $s$ in Fig. \ref{Contour}(a). In the calculation, we assume $G'=3$, $r=1$, and the quantum efficiencies of D1 and D2 are perfect. For each combination of $s$ and $p$, we compute the terms in Eqs. (\ref{eq:perfect_phase_matching_filtered_signal_ave_pho_num})-(\ref{Hsi2_f}) and substitute the results into Eq. (\ref{eq:R_t2}). Fig. \ref{Contour}(a) shows that $R_t$ depends on both $s$ and $p$. We notice that with the increase of $p$, the dependence of $R_t$ upon $s$ becomes stronger. When $p$ is less than $\sim 1.75$, at a fixed $p$, $R_t$ increases with $s$, but the noise reduction condition of $R_t<1$ is always achievable. However, when $p$ is greater than $\sim 1.75$, to obtain the noise reduction of $R_t<1$, the value of $s$ should be smaller than a certain value, which decreases with the increase of $p$. In real experiment, it is necessary to use the filter $F_{s(i)}$ with a certain bandwidth to prevent the strong pump background from reaching the detectors D1 and D2. So, to improve the noise reduction of twin beams, the pump with a pulse duration longer than that of the input signal, i.e., $p<1$, is desirable.

Figure \ref{Contour}(a) also shows that the contour $R_t=1$ divides the plot into two zones according to $R_t\leq 1$ or $R_t>1$. For the former case, there is a one-to-one correspondence between the combination of $s$, $p$, and $R_t$; while for the latter case, each $R_t$ maps two values of $s$ at a fixed value of $p$: one is very close to $0$, the other is larger. To better understand Fig. \ref{Contour}(a), we also plot the contours of the sums of the positive terms and negative terms in Eq. (\ref{eq:R_t2}) in Figs. \ref{Contour}(b) and \ref{Contour}(c), which respectively illustrate the influence of collection efficiency upon the intensity noise of individual beams and upon the correlation of twin beams. For both Figs. \ref{Contour}(b) and \ref{Contour}(c), the contour decreases with the increase of $s$ ($p$) at a fixed value of $p$ ($s$). However, since the correlation of twin beams highly relies on the collection efficiency, the dependence of the contour on $s$ and $p$ in Fig. \ref{Contour}(c) is stronger than that in Fig. \ref{Contour}(b). In particular, when $p$ increases, the value of $2(H_{si1}+H_{si2})|\alpha|^2/(I_s+I_i)$ in Fig. \ref{Contour}(c) rapidly decreases with the increase of $s$, which is responsible for the multivalued mappings in the zone of $R_t>1$ (see Fig. \ref{Contour}(a)).

\begin{figure}[]
 \includegraphics[width=0.66\textwidth]{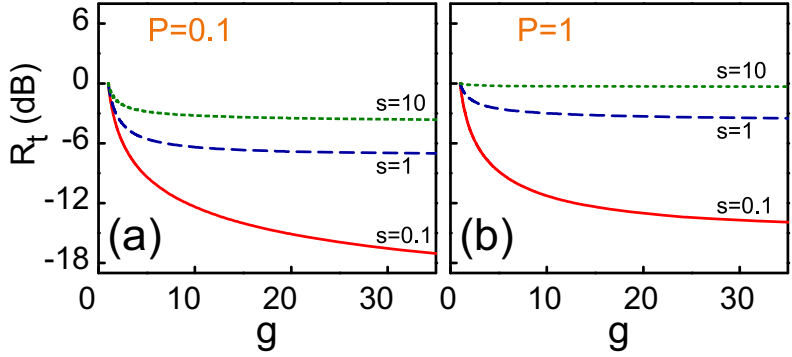}
 \caption{\label{fig:twin_beams_filter_bandwidth}The normalized intensity difference noise of the twin beams $R_t$ versus $g$ for (a) $p = 0.1$, (b) $p=1$, respectively. For each setting of $p$, $s$ in Eq.(\ref{eq:perfect_phase_matching_filtered_signal_ave_pho_num}) and (\ref{Hs1_f})-(\ref{Hvaci_f}) is $s=0.1$, $s=1$, and $s=10$, respectively. In this calculation, $\eta_s = \eta_i = 1$. }
 \end{figure}

To further demonstrate the dependence of the noise reduction of $R_t$ upon the photon number gain $g$, we plot $R_t$ as a function of $g$ for $s=0.1$, $s=1$ and $s=10$. In Figs. \ref{fig:twin_beams_filter_bandwidth}(a) and \ref{fig:twin_beams_filter_bandwidth}(b), we assume $\eta_s=\eta_i=1$ and $r=1$, and $p$ is fixed at $p=0.1$ and $p=1$, respectively. For each combination of $s$ and $p$, we compute the terms in Eqs. (\ref{eq:perfect_phase_matching_filtered_signal_ave_pho_num})-(\ref{Hvaci_f}) by changing $G'$ ($g$) and substitute the results into Eq.(\ref{eq:R_t2}). It is obvious that for $p$ and $s$ with fixed values, $R_t$ decreases with the increase of $g$. Moreover, at a fixed value of $g$ and $s$ ($p$), $R_t$ increases with the increase of $p$ ($s$) due to the decreased collection efficiency of twin beams.

\begin{figure}[]
 \includegraphics[width=0.66\textwidth]{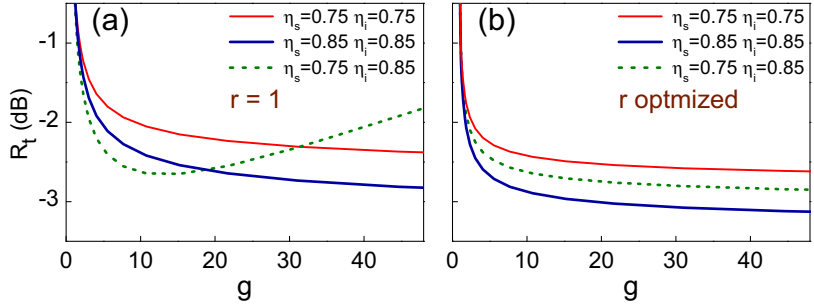}
 \caption{\label{fig:twin_beams_filtered_effeciency}The normalized intensity difference noise of the twin beams $R_t$ versus g for different setting of $\eta_s$ and $\eta_i$. For the plots in (a), $r = 1$; while for the plots in (b), r is optimized for minimizing the measured $R_t$. In this calculation, $p = 1$ and $s = 1$.}
 \end{figure}

Having understood the influence of the collection efficiency of twin beams on $R_t$, we then study the influence of the detection efficiency on $R_t$ when the values of $p$ and $s$ are fixed. Without loss of generality, we assume $p=1$ and $s=1$. When the AC response ratio is set to $r=1$, we plot $R_t$ as a function of $g$ by varying $\eta_s$ and $\eta_i$, as shown in Fig. \ref{fig:twin_beams_filtered_effeciency}(a). It is obvious that for the case of $\eta_s=\eta_i$, $R_t$ decreases with the increase of $\eta_{s(i)}$ at a fixed value of $g$. However, for the case of $\eta_s\neq\eta_i$, $R_t$ is not a monotonically decreasing function of $g$ and $\eta_{s(i)}$. Depending on $g$, the correlation of photon currents originated from the quantum correlation of the signal and idler beams might become weak or strong due to the unbalanced detection efficiencies. As shown in Fig. \ref{fig:twin_beams_filtered_effeciency}(a), $R_t$ for the case of  $\eta_s=75\%$ and $\eta_i=85\%$ is lower than that for $\eta_s=\eta_i=85\%$ when $g$ is less than 18; however, $R_t$ becomes even higher than that for $\eta_s=\eta_i=75\%$ when $g$ is greater than 30. Making comparison between Fig. \ref{fig:twin_beams_filtered_effeciency}(a) and Fig. 2(a), which are obtained for FOPA with non-factorable JSF and with single frequency pump, respectively, we find there are two differences: (i) for $g$, $\eta_s$ and $\eta_i$ with certain values, $R_t$ in Fig. \ref{fig:twin_beams_filtered_effeciency}(a) is larger than that in Fig. 2(a) because the collection efficiency of the pulsed twin beams is smaller; (ii) the difference between the curve for $\eta_s=75\%$ and $\eta_i=85\%$ and that for $\eta_s=\eta_i=85\%$ ($\eta_s=\eta_i=75\%$) in Fig. \ref{fig:twin_beams_filtered_effeciency}(a) is more apparent than that in Fig. 2(a). Therefore, for the pulse pumped FOPA, the measured $R_t$ can be minimized by properly adjusting the efficiency $\eta_s$ or $\eta_i$~\cite{Guo12}. However, this method will introduce extra loss, and the minimized $R_t$ can only be obtained for $g$ with a specified value.

If the ratio $r$ is adjustable, we can minimize $R_t$ by optimizing the value of $r$ without introducing the vacuum noise. In this case, as shown in Fig. \ref{fig:twin_beams_filtered_effeciency}(b), for fixed values of $\eta_s$ and $\eta_i$, $R_t$ always decreases with the increase of $g$; while for a fixed value of $g$, $R_t$ always decreases with the increase of $\eta_{s}$ and $\eta_{i}$. Hence, for both the signal and idler channels, the higher the efficiencies are, the better the noise reduction is.

\subsection{\label{input_signal_noise}Influence of excess noise of input signal}

Finally, we analyze the influence of the excess noise of input signal on $R_t$, since the noise of input signal pulses originated from the mode-locked fiber lasers is often higher than the SNL~\cite{Guo12}.
When the excess noise is included, the input signal can be viewed as a mixture of coherent states, whose density operator and average photon number are written as
\begin{equation}
\label{P_rep}
\hat{\rho_I} =   \int P( \alpha') | \alpha' \rangle \langle \alpha' | d^2\alpha' ~~
\end{equation}
and
\begin{equation}
\label{I0-mixer}
\overline{I}_0 = \int P(\alpha') |\alpha'|^2 d^2\alpha',
\end{equation}
respectively, where $\alpha'$ is a complex random variable and $P(\alpha')$ is the classical probability density.
Consequently,
the formulas of the average photon number of amplified signal and generated idler beams, $\overline{I}_s$ and $\overline{I}_i$, can be obtained by replacing the term $|\alpha|^2$ in Eq. (\ref{eq:signal_idler_Average Photon  Number}) with $\overline{I}_0$.

We then calculate the quantum noise $\Delta I_{s(i)}^2$ in Eq. (\ref{eq:fluctuation_general}) and $\Delta I_{t}^2$ in Eq. (\ref{difference-fluctuation_general}) by taking the quantum average over the density operator $\hat{\rho_I}$. After some algebra, we get
the normalized intensity noise of the amplified signal (generated idler) beam
\begin{equation}
\label{Rj-excess}
 R_{s(i)}'=R_{s(i)}+\overline{I}_{s(i)}\left(\langle|\alpha'|^4\rangle-(\langle|\alpha'|^2\rangle)^2\right),
\end{equation}
and the normalized intensity difference noise of twin beams
\begin{equation}
\label{Rt-excess}
R_{t}'=R_{t}+\frac{(\overline{I}_s-r\overline{I}_i)^2}{\overline{I}_s+r\overline{I}_i}\left(\langle|\alpha'|^4\rangle-(\langle|\alpha'|^2\rangle)^2\right),
\end{equation}
where $\langle|\alpha'|^n\rangle = \int P(\alpha') |\alpha'|^n d^2\alpha'$, and the inequality  $\langle|\alpha'|^4\rangle-(\langle|\alpha'|^2\rangle)^2 >0$ holds for the input signal with excess noise. In Eqs. (\ref{Rj-excess}) and (\ref{Rt-excess}), the first terms $R_{s(i)}$ and $R_{t}$ in the right hand sides are the corresponding noise for input signal in a pure coherent state with $I_{in}=\overline{I}_0$ (see Eqs. (\ref{RsAndRi}) and (\ref{eq:R_t2})); while the second terms in the right hand sides are originated from the excess noise of input signal.

%which adds more noise to the output fields of an FOPA and prevents the intensity difference noise of twin beams from further reduction.

Equations (\ref{Rj-excess}) and (\ref{Rt-excess}) indicate that the side effect of the excess noise can be eliminated under the condition $r=\overline{I}_s/\overline{I}_i$, and the experiment in Ref.~\cite{Guo12} has verified this for the case of $r=1$. In the low gain regime, the value of $r=\overline{I}_s/\overline{I}_i$ is different from $r_{opt}$ for minimizing $R_t$ (see Fig. \ref{fig:twin_beams_no_filtered}(b)), so the excess noise the input signal is deleterious for the noise reduction of $R_t$. However, in the high gain limit, we have $r=I_s/I_i\rightarrow 1$. In this case, eliminating the side effect of excess signal noise and minimizing $R_t$ can be achieved simultaneously. Therefore, a high gain FOPA is desirable for improving the noise reduction of twin beams.

\section{\label{Conclusion}Summary and Discussions:\protect\\}
In conclusion, We have developed the multi-mode quantum theory for analyzing the noise characteristics of a non-degenerate phase insensitive FOPA pumped by a pulsed laser. The calculation shows that both the noise figure and the intensity difference noise of twin beams depend on the JSF. In the high gain regime, the noise figure is generally greater than the 3dB quantum limit unless the JSF is factorable and the spectrum of the injected signal well matches the gain spectrum in signal band; while the intensity difference noise can be significantly less than the SNL. To closely resemble the experiments, the quantum noise of twin beams generated by a broadband FOPA is numerically studied by taking the real experimental conditions into account. In addition to the influences of the quantum efficiency of detectors and the excesses noise of input signal, the influence of the collection efficiency of the twin beams, determined by the ratio of the pump bandwidth to signal bandwidth $p$ and the ratio of the signal bandwidth to filter
bandwidth $s$, are carefully analyzed as well. Our theoretical investigation is of practical importance, it not only serves as a guideline for optimizing the measured noise reduction of the twin beams, but also helps for understanding the quantum noise preformance of pulse pumped FOPA.

Instead of using the Bloch-Messiah reduction or Schmidt decomposition of the JSF, which has been exploited to theoretically analyze the pulsed CV nonclassical light via the high gain spontaneous parametric down conversions in Refs.~\cite{Wasi06, Silberhorn2011}, our transformation functions of the Bogoliubov transformation are expressed in the infinite series of the gain coefficient $G$. The accuracy of our numerical simulation can be controlled by the two parameters: one is the precision of the JSF determined by the step size of the frequency range, the other is the order of the series determined by the gain of FWM. So our method has a greater flexibility.

We believe our theory can be extended to a more generalized OPA, including the degenerate OPA and the phase-sensitive OPA. Moreover, our theory can also be used to study the quantum noise of other quantities, such as the noise correlation between the quadrature amplitudes of the signal and idler beams. However, it is worth pointing out that the FOPA in our model is simplified as a linear amplifier, and we only considered the FWM process. Therefore, our calculation only quantitatively explains the results obtained under the condition of negligible pump depletion~\cite{Guo12}. To model the FOPA system more accurately, other effects like Raman effect, higher-order FWM and gain saturation effect should be included as well~\cite{Guo12}.

\begin{acknowledgments}
This work was supported in part by the State Key Development Program for Basic Research of China (No. 2010CB923101) and by the Specialized Research Fund for the Doctoral Program of Higher Education of China (No. 20120032110055).
\end{acknowledgments}


\begin{thebibliography}{31}
\expandafter\ifx\csname natexlab\endcsname\relax\def\natexlab#1{#1}\fi
\expandafter\ifx\csname bibnamefont\endcsname\relax
  \def\bibnamefont#1{#1}\fi
\expandafter\ifx\csname bibfnamefont\endcsname\relax
  \def\bibfnamefont#1{#1}\fi
\expandafter\ifx\csname citenamefont\endcsname\relax
  \def\citenamefont#1{#1}\fi
\expandafter\ifx\csname url\endcsname\relax
  \def\url#1{\texttt{#1}}\fi
\expandafter\ifx\csname urlprefix\endcsname\relax\def\urlprefix{URL }\fi
\providecommand{\bibinfo}[2]{#2}
\providecommand{\eprint}[2][]{\url{#2}}

\bibitem[{\citenamefont{Tong et~al.}(2011)\citenamefont{Tong, Lundstr\"{o}m,
  Andrekson, McKinstrie, Karlsson, Blessing, Tipsuwannakul, Puttnam, Toda, and
  Gr\"{u}ner-Nielsen}}]{Tong2011}
\bibinfo{author}{\bibfnamefont{Z.}~\bibnamefont{Tong}},
  \bibinfo{author}{\bibfnamefont{C.}~\bibnamefont{Lundstr\"{o}m}},
  \bibinfo{author}{\bibfnamefont{P.~A.} \bibnamefont{Andrekson}},
  \bibinfo{author}{\bibfnamefont{C.~J.} \bibnamefont{McKinstrie}},
  \bibinfo{author}{\bibfnamefont{M.}~\bibnamefont{Karlsson}},
  \bibinfo{author}{\bibfnamefont{D.~J.} \bibnamefont{Blessing}},
  \bibinfo{author}{\bibfnamefont{E.}~\bibnamefont{Tipsuwannakul}},
  \bibinfo{author}{\bibfnamefont{B.~J.} \bibnamefont{Puttnam}},
  \bibinfo{author}{\bibfnamefont{H.}~\bibnamefont{Toda}}, \bibnamefont{and}
  \bibinfo{author}{\bibfnamefont{L.}~\bibnamefont{Gr\"{u}ner-Nielsen}},
  \bibinfo{journal}{Nature Photonics} \textbf{\bibinfo{volume}{5}},
  \bibinfo{pages}{430} (\bibinfo{year}{2011}).

\bibitem[{\citenamefont{Hansryd et~al.}(2002)\citenamefont{Hansryd, Andrekson,
  Westlund, Li, and Hedekvist}}]{Hans02}
\bibinfo{author}{\bibfnamefont{J.}~\bibnamefont{Hansryd}},
  \bibinfo{author}{\bibfnamefont{P.~A.} \bibnamefont{Andrekson}},
  \bibinfo{author}{\bibfnamefont{M.}~\bibnamefont{Westlund}},
  \bibinfo{author}{\bibfnamefont{J.}~\bibnamefont{Li}}, \bibnamefont{and}
  \bibinfo{author}{\bibfnamefont{P.~O.} \bibnamefont{Hedekvist}},
  \bibinfo{journal}{IEEE J. Sel. Top. Quant.} \textbf{\bibinfo{volume}{8}},
  \bibinfo{pages}{506} (\bibinfo{year}{2002}).

\bibitem[{\citenamefont{Marhic}(2008)}]{Marhic}
\bibinfo{author}{\bibfnamefont{M.~E.} \bibnamefont{Marhic}},
  \emph{\bibinfo{title}{Fiber Optical Parametric Amplifiers, Oscillators and
  Related Devices}} (\bibinfo{publisher}{Cambridge university press},
  \bibinfo{year}{2008}).

\bibitem[{\citenamefont{Agrawal}(2008)}]{Agrawal08}
\bibinfo{author}{\bibfnamefont{G.~P.} \bibnamefont{Agrawal}},
  \emph{\bibinfo{title}{Application of Nonlinear fiber optics}}
  (\bibinfo{publisher}{Academic Press}, \bibinfo{year}{2008}).

\bibitem[{\citenamefont{Caves}(1982)}]{Caves1982A}
\bibinfo{author}{\bibfnamefont{C.~M.} \bibnamefont{Caves}},
  \bibinfo{journal}{Physical Review D} \textbf{\bibinfo{volume}{26}},
  \bibinfo{pages}{1817} (\bibinfo{year}{1982}).

\bibitem[{\citenamefont{Wu et~al.}(1986)\citenamefont{Wu, Kimble, Hall, and
  Wu}}]{wu86}
\bibinfo{author}{\bibfnamefont{L.~A.} \bibnamefont{Wu}},
  \bibinfo{author}{\bibfnamefont{H.~J.} \bibnamefont{Kimble}},
  \bibinfo{author}{\bibfnamefont{J.~L.} \bibnamefont{Hall}}, \bibnamefont{and}
  \bibinfo{author}{\bibfnamefont{H.}~\bibnamefont{Wu}}, \bibinfo{journal}{Phys.
  Rev. Lett.} \textbf{\bibinfo{volume}{57}}, \bibinfo{pages}{2520}
  (\bibinfo{year}{1986}).

\bibitem[{\citenamefont{Ayt\"{u}r and Kumar}(1990)}]{Ayt90}
\bibinfo{author}{\bibfnamefont{O.}~\bibnamefont{Ayt\"{u}r}} \bibnamefont{and}
  \bibinfo{author}{\bibfnamefont{P.}~\bibnamefont{Kumar}},
  \bibinfo{journal}{Phys. Rev. Lett.} \textbf{\bibinfo{volume}{65}},
  \bibinfo{pages}{1551} (\bibinfo{year}{1990}).

\bibitem[{\citenamefont{Reid et~al.}(2009)\citenamefont{Reid, Drummond, Bowen,
  Cavalcanti, Lam, Bachor, Andersen, and Leuchs}}]{Reid09}
\bibinfo{author}{\bibfnamefont{M.~D.} \bibnamefont{Reid}},
  \bibinfo{author}{\bibfnamefont{P.~D.} \bibnamefont{Drummond}},
  \bibinfo{author}{\bibfnamefont{W.~P.} \bibnamefont{Bowen}},
  \bibinfo{author}{\bibfnamefont{E.~G.} \bibnamefont{Cavalcanti}},
  \bibinfo{author}{\bibfnamefont{P.~K.} \bibnamefont{Lam}},
  \bibinfo{author}{\bibfnamefont{H.~A.} \bibnamefont{Bachor}},
  \bibinfo{author}{\bibfnamefont{U.~L.} \bibnamefont{Andersen}},
  \bibnamefont{and} \bibinfo{author}{\bibfnamefont{G.}~\bibnamefont{Leuchs}},
  \bibinfo{journal}{Rev. Modern Phys.} \textbf{\bibinfo{volume}{81}},
  \bibinfo{pages}{1727} (\bibinfo{year}{2009}).

\bibitem[{\citenamefont{Sharping et~al.}(2001)\citenamefont{Sharping,
  Fiorentino, and Kumar}}]{Sharping01}
\bibinfo{author}{\bibfnamefont{J.~E.} \bibnamefont{Sharping}},
  \bibinfo{author}{\bibfnamefont{M.}~\bibnamefont{Fiorentino}},
  \bibnamefont{and} \bibinfo{author}{\bibfnamefont{P.}~\bibnamefont{Kumar}},
  \bibinfo{journal}{Opt. Lett.} \textbf{\bibinfo{volume}{26}},
  \bibinfo{pages}{367} (\bibinfo{year}{2001}).

\bibitem[{\citenamefont{Voss et~al.}(2006)\citenamefont{Voss,
  K\"{o}pr\"{u}l\"{u}, and Kumar}}]{Voss06}
\bibinfo{author}{\bibfnamefont{P.~L.} \bibnamefont{Voss}},
  \bibinfo{author}{\bibfnamefont{K.~G.} \bibnamefont{K\"{o}pr\"{u}l\"{u}}},
  \bibnamefont{and} \bibinfo{author}{\bibfnamefont{P.}~\bibnamefont{Kumar}},
  \bibinfo{journal}{J. Opt. Soc. Am. B} \textbf{\bibinfo{volume}{23}},
  \bibinfo{pages}{598} (\bibinfo{year}{2006}).

\bibitem[{\citenamefont{McKinstrie and Gordon}(2012)}]{Mck12}
\bibinfo{author}{\bibfnamefont{C.~J.} \bibnamefont{McKinstrie}}
  \bibnamefont{and} \bibinfo{author}{\bibfnamefont{J.~P.}
  \bibnamefont{Gordon}}, \bibinfo{journal}{IEEE J. Sel. Top. Quantum Electron.}
  \textbf{\bibinfo{volume}{18}}, \bibinfo{pages}{958} (\bibinfo{year}{2012}).

\bibitem[{\citenamefont{Tittel and Weihs}(2001)}]{Tittel01}
\bibinfo{author}{\bibfnamefont{W.}~\bibnamefont{Tittel}} \bibnamefont{and}
  \bibinfo{author}{\bibfnamefont{G.}~\bibnamefont{Weihs}},
  \bibinfo{journal}{Quantum Information and Computation}
  \textbf{\bibinfo{volume}{1}}, \bibinfo{pages}{3} (\bibinfo{year}{2001}).

\bibitem[{\citenamefont{Braunstein and Loock}(2005)}]{Braun05}
\bibinfo{author}{\bibfnamefont{S.~L.} \bibnamefont{Braunstein}}
  \bibnamefont{and} \bibinfo{author}{\bibfnamefont{P.~V.} \bibnamefont{Loock}},
  \bibinfo{journal}{Rev. Modern Phys.} \textbf{\bibinfo{volume}{77}},
  \bibinfo{pages}{513} (\bibinfo{year}{2005}).

\bibitem[{\citenamefont{Fiorentino et~al.}(2002)\citenamefont{Fiorentino, Voss,
  Sharping, and Kumar}}]{Fiorentino02}
\bibinfo{author}{\bibfnamefont{M.}~\bibnamefont{Fiorentino}},
  \bibinfo{author}{\bibfnamefont{P.~L.} \bibnamefont{Voss}},
  \bibinfo{author}{\bibfnamefont{J.~E.} \bibnamefont{Sharping}},
  \bibnamefont{and} \bibinfo{author}{\bibfnamefont{P.}~\bibnamefont{Kumar}},
  \bibinfo{journal}{Photon. Technol. Lett.} \textbf{\bibinfo{volume}{14}},
  \bibinfo{pages}{983} (\bibinfo{year}{2002}).

\bibitem[{\citenamefont{Li et~al.}(2005)\citenamefont{Li, Voss, Sharping, and
  Kumar}}]{Li05a}
\bibinfo{author}{\bibfnamefont{X.}~\bibnamefont{Li}},
  \bibinfo{author}{\bibfnamefont{P.~L.} \bibnamefont{Voss}},
  \bibinfo{author}{\bibfnamefont{J.~E.} \bibnamefont{Sharping}},
  \bibnamefont{and} \bibinfo{author}{\bibfnamefont{P.}~\bibnamefont{Kumar}},
  \bibinfo{journal}{Phys. Rev. Lett.} \textbf{\bibinfo{volume}{94}},
  \bibinfo{pages}{053601} (\bibinfo{year}{2005}).

\bibitem[{\citenamefont{Takesue}(2006)}]{Takesue06}
\bibinfo{author}{\bibfnamefont{H.}~\bibnamefont{Takesue}},
  \bibinfo{journal}{Opt. Express} \textbf{\bibinfo{volume}{14}},
  \bibinfo{pages}{3453} (\bibinfo{year}{2006}).

\bibitem[{\citenamefont{Alibart et~al.}(2006)\citenamefont{Alibart, Fulconis,
  Wong, Murdoch, Wadsworth, and Rarity}}]{Alibart06}
\bibinfo{author}{\bibfnamefont{O.}~\bibnamefont{Alibart}},
  \bibinfo{author}{\bibfnamefont{J.}~\bibnamefont{Fulconis}},
  \bibinfo{author}{\bibfnamefont{G.~K.~L.} \bibnamefont{Wong}},
  \bibinfo{author}{\bibfnamefont{S.~G.} \bibnamefont{Murdoch}},
  \bibinfo{author}{\bibfnamefont{W.~J.} \bibnamefont{Wadsworth}},
  \bibnamefont{and} \bibinfo{author}{\bibfnamefont{J.~G.}
  \bibnamefont{Rarity}}, \bibinfo{journal}{New J. Phys.}
  \textbf{\bibinfo{volume}{8}}, \bibinfo{pages}{67} (\bibinfo{year}{2006}).

\bibitem[{\citenamefont{Goldschmidt et~al.}(2008)\citenamefont{Goldschmidt,
  Eisaman, Fan, Polyakov, and Migdall}}]{Fan08}
\bibinfo{author}{\bibfnamefont{E.~A.} \bibnamefont{Goldschmidt}},
  \bibinfo{author}{\bibfnamefont{M.~D.} \bibnamefont{Eisaman}},
  \bibinfo{author}{\bibfnamefont{J.}~\bibnamefont{Fan}},
  \bibinfo{author}{\bibfnamefont{S.~V.} \bibnamefont{Polyakov}},
  \bibnamefont{and} \bibinfo{author}{\bibfnamefont{A.}~\bibnamefont{Migdall}},
  \bibinfo{journal}{Phys. Rev. A} \textbf{\bibinfo{volume}{78}},
  \bibinfo{pages}{013844} (\bibinfo{year}{2008}).

\bibitem[{\citenamefont{Li et~al.}(2004)\citenamefont{Li, Chen, Voss, Sharping,
  and Kumar}}]{Li04}
\bibinfo{author}{\bibfnamefont{X.}~\bibnamefont{Li}},
  \bibinfo{author}{\bibfnamefont{J.}~\bibnamefont{Chen}},
  \bibinfo{author}{\bibfnamefont{P.~L.} \bibnamefont{Voss}},
  \bibinfo{author}{\bibfnamefont{J.}~\bibnamefont{Sharping}}, \bibnamefont{and}
  \bibinfo{author}{\bibfnamefont{P.}~\bibnamefont{Kumar}},
  \bibinfo{journal}{Opt. Express} \textbf{\bibinfo{volume}{12}},
  \bibinfo{pages}{3737} (\bibinfo{year}{2004}).

\bibitem[{\citenamefont{Guo et~al.}(2012)\citenamefont{Guo, Li, Liu, , Yang,
  and Ou}}]{Guo12}
\bibinfo{author}{\bibfnamefont{X.}~\bibnamefont{Guo}},
  \bibinfo{author}{\bibfnamefont{X.}~\bibnamefont{Li}},
  \bibinfo{author}{\bibfnamefont{N.}~\bibnamefont{Liu}}, ,
  \bibinfo{author}{\bibfnamefont{L.}~\bibnamefont{Yang}}, \bibnamefont{and}
  \bibinfo{author}{\bibfnamefont{Z.~Y.} \bibnamefont{Ou}},
  \bibinfo{journal}{Appl. Phys. Lett.} \textbf{\bibinfo{volume}{101}},
  \bibinfo{pages}{261111} (\bibinfo{year}{2012}).

\bibitem[{\citenamefont{Laurat et~al.}(2003)\citenamefont{Laurat, Coudreau,
  Treps, Ma\^{\i}tre, and Fabre}}]{Lau03}
\bibinfo{author}{\bibfnamefont{J.}~\bibnamefont{Laurat}},
  \bibinfo{author}{\bibfnamefont{T.}~\bibnamefont{Coudreau}},
  \bibinfo{author}{\bibfnamefont{N.}~\bibnamefont{Treps}},
  \bibinfo{author}{\bibfnamefont{A.}~\bibnamefont{Ma\^{\i}tre}},
  \bibnamefont{and} \bibinfo{author}{\bibfnamefont{C.}~\bibnamefont{Fabre}},
  \bibinfo{journal}{Phys. Rev. Lett.} \textbf{\bibinfo{volume}{91}},
  \bibinfo{pages}{213601} (\bibinfo{year}{2003}).

\bibitem[{\citenamefont{Zhang et~al.}(2007)\citenamefont{Zhang, Furuta, Okubo,
  Takahashi, and Hirano}}]{Zhang07}
\bibinfo{author}{\bibfnamefont{Y.}~\bibnamefont{Zhang}},
  \bibinfo{author}{\bibfnamefont{T.}~\bibnamefont{Furuta}},
  \bibinfo{author}{\bibfnamefont{R.}~\bibnamefont{Okubo}},
  \bibinfo{author}{\bibfnamefont{K.}~\bibnamefont{Takahashi}},
  \bibnamefont{and} \bibinfo{author}{\bibfnamefont{T.}~\bibnamefont{Hirano}},
  \bibinfo{journal}{Phys. Rev. A} \textbf{\bibinfo{volume}{76}},
  \bibinfo{pages}{012314} (\bibinfo{year}{2007}).

\bibitem[{\citenamefont{Wasilewski et~al.}(2006)\citenamefont{Wasilewski,
  Lvovsky, Banaszek, and Radzewicz}}]{Wasi06}
\bibinfo{author}{\bibfnamefont{W.}~\bibnamefont{Wasilewski}},
  \bibinfo{author}{\bibfnamefont{A.~I.} \bibnamefont{Lvovsky}},
  \bibinfo{author}{\bibfnamefont{K.}~\bibnamefont{Banaszek}}, \bibnamefont{and}
  \bibinfo{author}{\bibfnamefont{C.}~\bibnamefont{Radzewicz}},
  \bibinfo{journal}{Phys. Rev. A} \textbf{\bibinfo{volume}{73}},
  \bibinfo{pages}{063819} (\bibinfo{year}{2006}).

\bibitem[{\citenamefont{Christ et~al.}(2011)\citenamefont{Christ, Laiho,
  Eckstein, Cassemiro, and Silberhorn}}]{Silberhorn2011}
\bibinfo{author}{\bibfnamefont{A.}~\bibnamefont{Christ}},
  \bibinfo{author}{\bibfnamefont{K.}~\bibnamefont{Laiho}},
  \bibinfo{author}{\bibfnamefont{A.}~\bibnamefont{Eckstein}},
  \bibinfo{author}{\bibfnamefont{K.~N.} \bibnamefont{Cassemiro}},
  \bibnamefont{and}
  \bibinfo{author}{\bibfnamefont{C.}~\bibnamefont{Silberhorn}},
  \bibinfo{journal}{New J. Phys.} \textbf{\bibinfo{volume}{13}},
  \bibinfo{pages}{033027} (\bibinfo{year}{2011}).

\bibitem[{\citenamefont{Ou and Kimble}(1995)}]{Ou1995}
\bibinfo{author}{\bibfnamefont{Z.~Y.} \bibnamefont{Ou}} \bibnamefont{and}
  \bibinfo{author}{\bibfnamefont{H.~J.} \bibnamefont{Kimble}},
  \bibinfo{journal}{Phys. Rev. A} \textbf{\bibinfo{volume}{52}},
  \bibinfo{pages}{3126} (\bibinfo{year}{1995}).

\bibitem[{\citenamefont{Stolen and Bjorkholm}(1982)}]{Stolen82}
\bibinfo{author}{\bibfnamefont{R.}~\bibnamefont{Stolen}} \bibnamefont{and}
  \bibinfo{author}{\bibfnamefont{J.~E.} \bibnamefont{Bjorkholm}},
  \bibinfo{journal}{J. Quantum Electron.} \textbf{\bibinfo{volume}{18}},
  \bibinfo{pages}{1062} (\bibinfo{year}{1982}).

\bibitem[{\citenamefont{Chen et~al.}(2005)\citenamefont{Chen, Li, and
  Kumar}}]{Chen05}
\bibinfo{author}{\bibfnamefont{J.}~\bibnamefont{Chen}},
  \bibinfo{author}{\bibfnamefont{X.}~\bibnamefont{Li}}, \bibnamefont{and}
  \bibinfo{author}{\bibfnamefont{P.}~\bibnamefont{Kumar}},
  \bibinfo{journal}{Physical Review A} \textbf{\bibinfo{volume}{72}},
  \bibinfo{pages}{033801} (\bibinfo{year}{2005}).

\bibitem[{\citenamefont{Yang et~al.}(2011)\citenamefont{Yang, Ma, Guo, Cui, and
  Li}}]{Yang2011}
\bibinfo{author}{\bibfnamefont{L.}~\bibnamefont{Yang}},
  \bibinfo{author}{\bibfnamefont{X.}~\bibnamefont{Ma}},
  \bibinfo{author}{\bibfnamefont{X.}~\bibnamefont{Guo}},
  \bibinfo{author}{\bibfnamefont{L.}~\bibnamefont{Cui}}, \bibnamefont{and}
  \bibinfo{author}{\bibfnamefont{X.}~\bibnamefont{Li}}, \bibinfo{journal}{Phys.
  Rev. A} \textbf{\bibinfo{volume}{83}}, \bibinfo{pages}{053843}
  (\bibinfo{year}{2011}).

\bibitem[{\citenamefont{Garay-Palmett et~al.}(2007)\citenamefont{Garay-Palmett,
  McGuinness, Cohen, Lundeen, Rangel-Rojo, U'Ren, Raymer, McKinstrie, Radic,
  and Walmsley}}]{Palmett07}
\bibinfo{author}{\bibfnamefont{K.}~\bibnamefont{Garay-Palmett}},
  \bibinfo{author}{\bibfnamefont{H.~J.} \bibnamefont{McGuinness}},
  \bibinfo{author}{\bibfnamefont{O.}~\bibnamefont{Cohen}},
  \bibinfo{author}{\bibfnamefont{J.~S.} \bibnamefont{Lundeen}},
  \bibinfo{author}{\bibfnamefont{R.}~\bibnamefont{Rangel-Rojo}},
  \bibinfo{author}{\bibfnamefont{A.~B.} \bibnamefont{U'Ren}},
  \bibinfo{author}{\bibfnamefont{M.~G.} \bibnamefont{Raymer}},
  \bibinfo{author}{\bibfnamefont{C.~J.} \bibnamefont{McKinstrie}},
  \bibinfo{author}{\bibfnamefont{S.}~\bibnamefont{Radic}}, \bibnamefont{and}
  \bibinfo{author}{\bibfnamefont{I.~A.} \bibnamefont{Walmsley}},
  \bibinfo{journal}{Opt. Express} \textbf{\bibinfo{volume}{15}},
  \bibinfo{pages}{14870} (\bibinfo{year}{2007}).

\bibitem[{\citenamefont{Cui et~al.}(2012)\citenamefont{Cui, Li, and
  Zhao}}]{Cui2012}
\bibinfo{author}{\bibfnamefont{L.}~\bibnamefont{Cui}},
  \bibinfo{author}{\bibfnamefont{X.}~\bibnamefont{Li}}, \bibnamefont{and}
  \bibinfo{author}{\bibfnamefont{N.}~\bibnamefont{Zhao}}, \bibinfo{journal}{New
  J. Phys.} \textbf{\bibinfo{volume}{14}}, \bibinfo{pages}{123001}
  (\bibinfo{year}{2012}).

\bibitem[{\citenamefont{Li et~al.}(2010)\citenamefont{Li, Ma, Quan, Yang, Cui,
  and Guo}}]{Li2010}
\bibinfo{author}{\bibfnamefont{X.}~\bibnamefont{Li}},
  \bibinfo{author}{\bibfnamefont{X.}~\bibnamefont{Ma}},
  \bibinfo{author}{\bibfnamefont{L.}~\bibnamefont{Quan}},
  \bibinfo{author}{\bibfnamefont{L.}~\bibnamefont{Yang}},
  \bibinfo{author}{\bibfnamefont{L.}~\bibnamefont{Cui}}, \bibnamefont{and}
  \bibinfo{author}{\bibfnamefont{X.}~\bibnamefont{Guo}}, \bibinfo{journal}{J.
  Opt. Soc. Am. B} \textbf{\bibinfo{volume}{27}}, \bibinfo{pages}{1857}
  (\bibinfo{year}{2010}).

\end{thebibliography}
 \end{document}